\documentclass[aps,prb,twocolumn,floatfix,superscriptaddress,preprintnumbers,showpacs]{revtex4-2}
\usepackage[T1]{fontenc}
\usepackage{verbatim}
\usepackage{amsmath}
\usepackage{amssymb}
\usepackage{graphicx,graphics}
\usepackage{hyperref}
\usepackage{float}
\usepackage{tikz}
\usepackage{tikz-3dplot}
\usetikzlibrary{positioning}
\usetikzlibrary{calc}
\usetikzlibrary{math}
\usepackage{calc}
\usepackage{xcolor}

\newcommand{\bx}{\textbf{x}}

\usetikzlibrary{arrows, decorations.markings}
\usetikzlibrary { decorations.pathmorphing, decorations.pathreplacing, decorations.shapes} 
\usetikzlibrary{decorations.pathreplacing}

\definecolor{bluenice}{RGB}{106, 174, 214}
\definecolor{rednice}{RGB}{250, 105, 73}
\definecolor{bordeau}{RGB}{167, 1, 69}

\tikzset{
    mark position/.style args={#1(#2)}{
        postaction={
            decorate,
            decoration={
                markings,
                mark=at position #1 with \coordinate (#2);
            }
        }
    }
}

\begin{document}
\title{Measurement-induced phase transition in a single-body tight-binding
model}
\author{Tony Jin}
\affiliation{Pritzker School of Molecular Engineering, The University of Chicago,
Chicago, IL 60637, USA}
\affiliation{Universit\'e C\^ote-d’Azur, CNRS, Centrale Med, Institut de Physique de Nice, 06200 Nice, France}\author{David G. Martin}
\affiliation{Kadanoff Center for Theoretical Physics and
Enrico Fermi Institute, 933 E 56th St, The University of Chicago, Chicago, IL 60637, USA}
\begin{abstract}
\noindent We study the statistical properties of a single free quantum particle
evolving coherently on a discrete lattice in ${\rm d}$ spatial dimensions
where every lattice site is additionally subject to continuous measurement of the
occupation number. Our numerical results indicate that the system undergoes a Measurement-induced Phase Transition
(MiPT) for ${\rm d}>1$ from a \textit{delocalized} to a \textit{localized} phase as the
measurement strength $\gamma$ is increased beyond a critical value
$\gamma_{c}$. In the language of surface growth, the delocalized phase corresponds to a \textit{smooth} phase while the localized phase corresponds to a \textit{rough} phase. We support our numerical results with perturbative renormalization group (RG) computations which are in qualitative agreement at one-loop order.
% \the\textwidth
\end{abstract}
\maketitle
Recently, it has been discovered that quantum chaotic systems subject
to continuous or projective measurements could undergo a phase transition
characterized by a change of the scaling properties of the entanglement
entropy with time or system size, a phenomenon now referred to as
Measurement-induced Phase Transition (MiPT) \citep{NahumMeasurementinducedtransition2019}.
MiPT constitutes a fascinating problem at the crossroad of statistical
physics and quantum information. 
As such, it has attracted a tremendeous
amount of interest in the recent years \citep{MeasurementinducedSchomerus,MeasurementinducedFisher,TurkeShiMiPThybrid,HuseMeasurementinduced,HuseMeasurementinduced3,HuseMeasurementinduced2,MeasurementinducedAltman,TurkeshiMiPTinfinitezeroclick,weinstein2022measurement,MeasurementBosonsRabl,YouennXhekSchiroMIPTnonHermitian,GranetMiPTPostSelection,GideonEEBKC}.
MiPTs are often characterized by a transition from
an \emph{area law }phase, \textit{i.e} a phase where the entanglement entropy
(EE) of a subsystem doesn't scale with its size, to a \emph{volume
law} phase, \textit{i.e} a phase where the EE scales with the system size.
Such a scaling transition occurs upon increasing the strength of the measurement and is observed in various systems such as in 1d interacting chaotic many-body systems. 

\begin{figure}[h]
\includegraphics[width=\columnwidth]{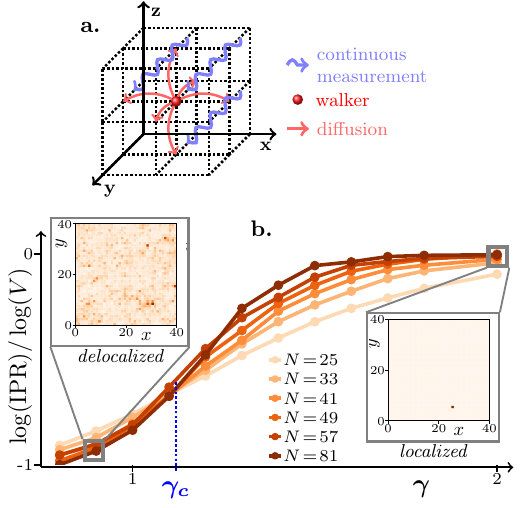}
\caption{\textbf{a.} A single quantum walker on a ${\rm d}$-dimensional square lattice undergoing a unitary tight-binding evolution and subject to independent continuous measurements of the occupation number at
every site. 
In ${\rm d}>1$, a phase transition from a
smooth/delocalized phase to a rough/localized phase occurs upon increasing the measurement rate $\gamma$ beyond a critical value $\gamma_{c}$. This is illustrated in
\textbf{b.} which shows the scaling of the Inverse Partition Ratio (defined as ${\rm IPR}=\sum_{\boldsymbol{j}}|\psi_{\boldsymbol{j}}|^4$) with the volume as a function of $\gamma$ for different system sizes in dimension $3$. In the delocalized phase, the IPR $\propto V^-1$ while the IPR $\propto 1$ in the localized phase.
%The delocalized phase is characterized by an IPR inversely proportional to $V$ while the localized phase exhibits a volume-independent IPR.
The insets show a typical density profile in each of these phases, where one of the spatial direction has been projected out. {\bf Parameters:} $\tau=1.5$, $N=41$, $dt=0.01$.} 
% For the insets$\gamma=0.9$  and for the rough phase $\gamma=2$. 

\label{fig:Figure-model+snapshots}
% \vspace{-0.8cm}
\end{figure}
However, surprisingly, the existence of a MiPT between two non-trivially entangled phases for free or Gaussian fermions undergoing measurements remains an actively debated question. 
% However, surprisingly, it turns out that free or Gaussian fermions or spins undergoing measurements escape this phenomenology and one of the actively debated question is the existence or lack thereof of a non-trivially entangled phase for these systems. 
While the original study of entanglement in 1d free fermions \citep{caoEntanglementFermionChain2019} showed
no signs of a phase transition, more recent numerical and theoretical investigations showed either the existence of a phase where the EE scales as $\log L$ \citep{BuchholdMiPTPRL,BuchholdMiPTPRX} for fermions with $U(1)$ symmetry or $\log L$ \citep{LudwigNLsM} and $\log^2 L$ \citep{NahumNLsM} for $\mathbb{Z}_2$ symmetry. Another recent study \citep{MirlinNLsM} also argued that the observed transitions for $U(1)$ fermions are in fact sharp crossovers. See also \cite{PRBPurificationtimescale} for a discussion of the roles played by global symmetries on purification times scales. 

In this work, we provide new insights on this conundrum by studying
the simpler, yet non trivial single-body problem of a particle
evolving coherently on a discrete lattice in ${\rm d}$ spatial dimensions,
where every lattice site is subject to independent, continuous measurements
of its occupation number--See Fig.~\ref{fig:Figure-model+snapshots}.
Combining numerical
simulations and perturbative renormalization group (RG) methods, we show that, while
we do not find evidence of a transition in ${\rm d}=1$,
there exists a phase transition from a \emph{smooth/delocalized} phase
to a \emph{rough/localized }phase when ${\rm d}>1$ (throughout the manuscript, a localized behavior refers to a spatially peaked wave function, without any statement about transport properties). Interestingly, this
shows that many-body effects are not necessary to observe a MiPT and corroborates
the result obtained in \citep{JinMartinMiPTclassical} for a classical
random walker undergoing continuous measurements. 

% \paragraph{Model}
\textit{Model}
We consider a single quantum particle on a square lattice of $V=N^{{\rm d}}$
sites with periodic boundary conditions. Let $\{\left|\boldsymbol{j}\right\rangle \}_{\boldsymbol{j}\in[1,N]^{d}}$
denote the position basis. The dynamics is described by a unitary
tight-binding evolution $\hat{H}:=-\tau\sum_{\{|\boldsymbol{e}|=1\}}\left|\boldsymbol{j}\right\rangle \left\langle \boldsymbol{j}+\boldsymbol{e}\right|$ where $\{|\boldsymbol{e}|=1\}$ is the set of vectors
of norm $1$.
%where $$ the subscript $(\boldsymbol{j},\boldsymbol{k})$ denotes adjacent sites.
In addition, each site undergoes continuous measurements of strength
$\gamma$ of the local occupation $\hat{n}_{\boldsymbol{j}}:=\left|\boldsymbol{j}\right\rangle \left\langle \boldsymbol{j}\right|$ resulting in the stochastic differential equation (SDE) \citep{JacobsintroductiontoCmeasurement}: 
\begin{align}
 & d\left|\psi\right\rangle =-iH\left|\psi\right\rangle dt\label{eq:fundequation}\\
 & +\sum_{\boldsymbol{j}}\left(-\frac{\gamma}{2}(\hat{n}_{\boldsymbol{j}}-\langle\hat{n}_{\boldsymbol{j}}\rangle_{t})^{2}dt+\sqrt{\gamma}(\hat{n}_{\boldsymbol{j}}-\langle\hat{n}_{\boldsymbol{j}}\rangle_{t})dB_{t}^{\boldsymbol{j}}\right)\left|\psi\right\rangle\;, \nonumber 
\end{align}
where $\langle\bullet\rangle_{t}:={\rm tr}(\rho_{t}\bullet)$. 
In \eqref{eq:fundequation}, the $\{B_{t}^{\boldsymbol{j}}\}_{\boldsymbol{j}\in[1,N]^{{\rm d}}}$ are
$N^{{\rm d}}$ independent Brownian processes with average $\mathbb{E}[dB_{t}^{\boldsymbol{j}}]=0$ and It\=o rules $dB_{t}^{\boldsymbol{j}}dB_{t'}^{\boldsymbol{k}}=\boldsymbol{1}_{0}(t-t')\delta_{\boldsymbol{j},\boldsymbol{k}}dt$
where $\boldsymbol{1}_{0}$ is the indicator function. This model was originally
introduced in \citep{BernardJinShpielberg_2018} for the free fermionic
case and has been subsequently studied in \citep{caoEntanglementFermionChain2019,BuchholdMiPTPRL,BuchholdMiPTPRX,Oded1Dmeasurement}
in the context of MiPTs -- see also \citep{JinTransportMeasures,FerreiraThermalenginesMeasure}
for applications to transport and thermal engines. 

In terms of the basis elements $\psi_{\boldsymbol{j}}$ defined as $\left|\psi\right\rangle =\sum_{\boldsymbol{j}}\psi_{\boldsymbol{j}}\left|\boldsymbol{j}\right\rangle $,
Eq.(\ref{eq:fundequation}) can be written as %\begin{widetext}

\begin{align}
d\psi_{\boldsymbol{j}}= & i\tau\sum_{\{|\boldsymbol{e}|=1\}}\psi_{\boldsymbol{j}+\boldsymbol{e}}dt-\frac{\gamma}{2}\psi_{\boldsymbol{j}}\big(1-2|\psi_{\boldsymbol{j}}|^{2}+\sum_{\boldsymbol{m}}|\psi_{\boldsymbol{m}}|^{4}\big)dt \nonumber\\
&+\sqrt{\gamma}\psi_{\boldsymbol{j}}\big(dB_{t}^{\boldsymbol{j}}-\sum_{\boldsymbol{m}}|\psi_{\boldsymbol{m}}|^{2}dB_{t}^{\boldsymbol{m}}\big).\label{eq:fundamentalequationpositionbasis}
\end{align}
%\end{widetext} 
Throughout the rest of the manuscript, we will fix the
initial condition of the system to be $\psi_{\boldsymbol{j}}(t=0)=N^{-{\rm d}/2}$ for all
$\boldsymbol{j}$. Note that by construction, $\sum_{\boldsymbol{j}}|\psi_{\boldsymbol{j}}|^{2}$
is preserved for each realization of the noise.

\begin{figure}
\includegraphics[width=\columnwidth]{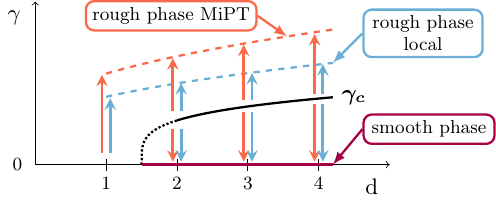}
\caption{Schematic of the renormalization flow of \eqref{eq:localdiscreteequation} (blue) and \eqref{eq:fundamentalequationpositionbasis} (red) as a function of dimension. 
Despite flowing toward different fixed points (\textit{i.e.} distinct critical exponents), both systems exhibit a similar phenomenology: when ${\rm d}\geq 2$ they feature a genuine phase transition between a smooth and a rough phase upon varying $\gamma$ while when ${\rm d}=1$ they only exhibit a single rough phase.}
\label{fig:transition_schematic}
\end{figure}

\begin{figure*}
\includegraphics[width=\textwidth]{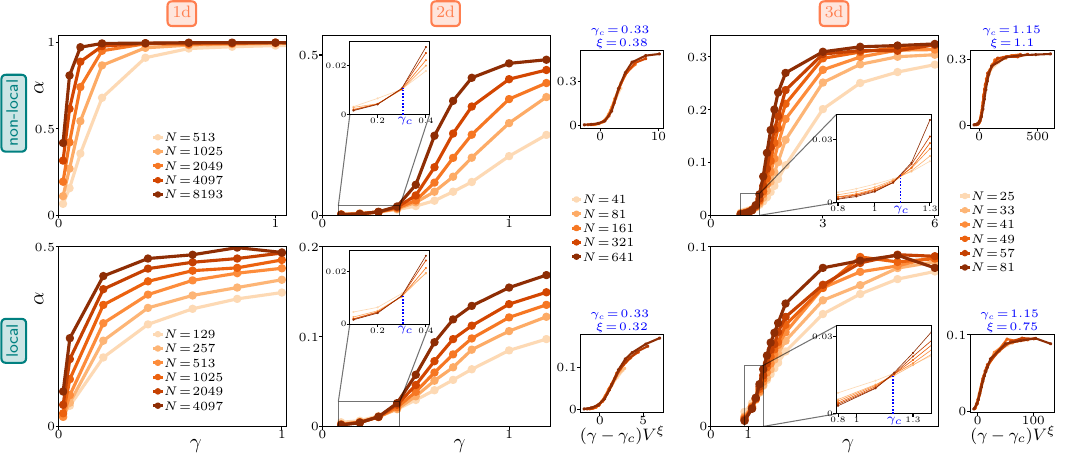}\caption{Critical exponent $\alpha$ as a function of the measurement strength
$\gamma$ obtained in numerical simulations of \eqref{eq:fundamentalequationpositionbasis} (non-local) and \eqref{eq:localdiscreteequation} (local) in dimension $1$, $2$ and $3$ (from left to right) for different system sizes $N$.
In both $2{\rm d}$ and $3{\rm d}$ we observe a transition from a smooth to a rough phase: the insets are zoom over the critical $\gamma_c$ at which these transitions occur. 
Smaller plots show the finite-size scaling $\alpha=f\left((\gamma-\gamma_c)V^{\xi}\right)$ corresponding to the graphs on their left.
{\bf Parameters:} $\tau=1$, $dt=0.01$.}
\label{fig:Critical-exponent-IPR}
\end{figure*}

Even though (\ref{eq:fundamentalequationpositionbasis}) describes
the dynamics of a single particle, getting an exact solution of such a
SDE is in general a formidable
task. One way to make progress is to restrict it to what
we will refer to as the \emph{delocalized} phase, \textit{i.e} to assume that
$|\psi_{\boldsymbol{j}}|$ is of order $N^{-{\rm d}/2}$. Under this assumption,
keeping the leading order in $N^{-1}$ in (\ref{eq:fundamentalequationpositionbasis})
gives the simpler expression 
\begin{equation}
d\psi_{\boldsymbol{j}}=\bigg(i\tau\sum_{\{|\boldsymbol{e}|=1\}}\psi_{\boldsymbol{j}+\boldsymbol{e}}-\frac{\gamma}{2}\psi_{\boldsymbol{j}}\bigg)dt+\sqrt{\gamma}\psi_{\boldsymbol{j}}dB_{t}^{\boldsymbol{j}}\label{eq:localdiscreteequation}
\end{equation}
which is now \emph{local} and \emph{linear} in $\psi_{\boldsymbol{j}}$. We further take the continuous limit by
introducing the lattice spacing $b$ and the continuous quantities
$\vec{r}=\boldsymbol{j}b$, $\varphi(\vec{r}=\boldsymbol{j}b)=b^{-{\rm d}/2}\psi_{\boldsymbol{j}}$,
$d\eta(\vec{r},t):=b^{-{\rm d}/2}dB_{t}^{\boldsymbol{j}}$, $D:=b^{2}\tau$,
$\lambda:=\gamma b^{{\rm d}}$. Up to a global phase, \eqref{eq:localdiscreteequation} then becomes 
\begin{equation}
d\varphi=\left(iD\nabla^{2}\varphi-\frac{\gamma}{2}\varphi\right)dt+\sqrt{\lambda}\varphi d\eta.\label{eq:localcontinuousequation}
\end{equation}
It is important to note that the noise becomes multiplicative in \eqref{eq:localcontinuousequation}.
This allows us to draw an analogy between \eqref{eq:localcontinuousequation} and the Stochastic Heat Equation (SHE), thereby relating \eqref{eq:localdiscreteequation} to KPZ physics \citep{KPZpaper,BertiniSHEFeynmanKac}.
Such an analogy was already fruitfully exploited in \citep{JinMartinMiPTclassical}, where it led to an intuitive understanding of a MiPT in a classical context.
The difference with this previous study is that we deal with an \emph{imaginary} diffusion term $D$ as well as a real ``mass'' term $\gamma/2$. We will see that, perhaps surprisingly, this qualitatively modifies the phase diagram as shown in Fig.~\ref{fig:transition_schematic}. 
Most notably, the \textit{lower critical dimension becomes 1 instead of 2}. We observe this feature numerically, both for the local \eqref{eq:localdiscreteequation} and non-local model \eqref{eq:fundamentalequationpositionbasis} (see Fig.~\ref{fig:Critical-exponent-IPR}).

%Even though these differences lead to quantitative modifications with respect to the SHE, we will see that one of its main feature, namely the existence of a phase transition with lower critical dimension $2$, remains present in  \eqref{eq:localcontinuousequation}. 

% Anticipating a bit, for the purpose of studying phase transition,
We expect Eq.~\eqref{eq:localdiscreteequation} to be valid as long as $|\psi_{\boldsymbol{j}}|$ remains close to the homogeneous profile of order $N^{-{\rm d}/2}$. 
Such an assumption is verified when the renormalization flow is directed towards the delocalized phase: in this case, $|\psi_{\boldsymbol{j}}|$ is indeed driven closer to the homogeneous profile. 
On the opposite, if the renormalization flow is directed towards the localized phase, $|\psi_{\boldsymbol{j}}|$ is driven away from the homogeneous profile and we don't expect \eqref{eq:localdiscreteequation} to remain valid at long time. 
Since we start from a flat initial condition, we also expect the local approximation to be good at short times, i.e for $t\ll\frac{1}{\gamma}$.
Although (\ref{eq:localdiscreteequation}) does not describe the strongly localized regime, it still allows us to study the critical line separating the two phases. Note that this approach was successfully used in a previous work to study the classical counterpart of \eqref{eq:fundamentalequationpositionbasis} \citep{JinMartinMiPTclassical}.

\begin{figure}
\begin{comment}
\begin{tikzpicture}
\tikzmath{\ylab=1.5;\scalelab=1.2;}
\node[font=\bf,scale=\scalelab] at (-3.35,\ylab) {a.};
\node[font=\bf,scale=\scalelab] at (1.18,\ylab) {b.};
\node at (0,0) {\includegraphics[width=\columnwidth]{plot_growth_3D.pdf}};
\end{tikzpicture}
\end{comment}
\includegraphics[width=\columnwidth]{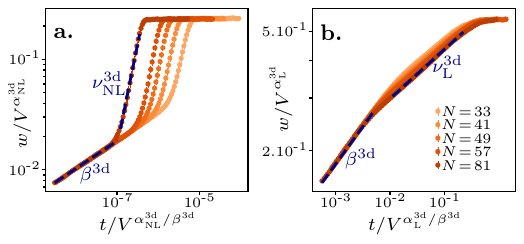}

\caption{
Log-Log plot of the rescaled width as a function of the rescaled time in dimension 3 for \eqref{eq:fundamentalequationpositionbasis} (\textbf{a.}) and \eqref{eq:localdiscreteequation} (\textbf{b.}).
Exponents: $\alpha^{3{\rm d}}_{\rm{NL}}=0.33$, $\nu^{3{\rm d}}_{\rm{NL}}=1.83$, $\alpha^{3{\rm d}}_{\rm{L}}=0.09$, $\nu^{3{\rm d}}_{\rm{L}}=0.16$ and $\beta^{3 {\rm d}}=0.25$.
\textbf{Parameters}: $\tau=1$, $dt=0.01$, $\gamma=8$.
}
\label{fig:growth_3D}
\end{figure}

% \paragraph{Numerical results}
\textit{Numerical results}
We begin by providing numerical simulations of the  microscopic non-local evolution \eqref{eq:fundamentalequationpositionbasis} (labeled NL)
and systematically compare with the approximate local equation \eqref{eq:localdiscreteequation} (labeled L). In order to characterize the different phases, we introduce the height $h_j$
\begin{equation}
h_{\boldsymbol{j}}:=\frac{1}{\sqrt{\gamma}}\log\left(|\psi_{\boldsymbol{j}}|^2\right).
\end{equation}
Drawing on the analogy with the classical counterpart of \eqref{eq:fundamentalequationpositionbasis} \cite{JinMartinMiPTclassical}, we expect that the width $w$ will follow a Family-Vicsek \cite{FractalconceptsBarbarasi,FamilyVicsek} type scaling according to 
\begin{equation}
w:=\sqrt{\frac{1}{V}\sum_{\boldsymbol{j}} (h_{\boldsymbol{j}}-\langle h\rangle_{\rm s})^2}\propto V^\alpha f\left(\frac{t}{V^{\alpha/\beta}}\right),
\label{eq:family_vicsek}
\end{equation}
where the bracket denotes the spatial average $\langle h\rangle_{\rm s}=\frac{1}{V}\sum_{\boldsymbol{j}} h_{\boldsymbol{j}}$. 
The function $f$ is such that $f(x)\propto x^\beta$ for $x\ll1$ while $f(x)\propto 1$ for $x\gg 1$. The universal exponents $\alpha$ and $\beta$ characterize the dynamical phases and are respectively called the \textit{roughness} and \textit{growth} exponents.

We show on Fig.~\ref{fig:growth_3D} a typical plot of the width as a function of time for both the local \eqref{eq:localdiscreteequation} and non-local evolution \eqref{eq:fundamentalequationpositionbasis} taken in the rough phase in ${\rm d}=3$. 
Note that the early time growth behavior are the same in both cases, indicating that the non-local terms does not contribute at this time scale. However, after this early regime, there is a crossover to another universality class, different for the local and non-local equations and characterized by the growth exponent $\nu_{\rm L/NL}$. See e.g. \citep{CrossoverKPZForrest,CrossoverKPZChame} for a discussion of crossovers in the real KPZ.

We further show on Fig.~\ref{fig:Critical-exponent-IPR}
the dependence of $\alpha$ as a function of $\gamma$ in ${\rm d}=1,2,3$ for different system sizes. In 1d, we see that all the curves collapse to the value $\alpha\approx1$ indicating a rough/localized phase while in {2d and} 3d, we see a clear crossing of the curves and a finite-size scaling collapse for both the local and non-local models at a finite
value of $\gamma=\gamma_c$.
This indicates a phase transition from a smooth/delocalized phase with $\alpha \approx 0$ to a rough/localized phase with finite $\alpha$. 
Importantly, while the local and non-local evolutions lead to different dynamical exponent in the strong coupling phase, they both entail the \textit{same value} for the critical rate $\gamma_c$ in ${\rm d}=2$ and ${\rm d}=3$. This confirms that the local equation is a good approximation in the smooth phase and is further able to capture the phase transition.

The numerically extracted values for $\alpha$, $\beta$ and $\nu$ in the rough phase are reported in table \ref{table:roughness}. 
Interestingly, we note that the value of $\alpha_{\rm L}$ only matches with the known values of the real KPZ in ${\rm d}=1$ \citep{NumericsKPZRSOSsurface}, further confirming that the strong coupling phase of the complex SHE lies in a different universality class than the real one when ${\rm d}>1$.

\begin{table}
\begin{center}
\begin{tabular}{ |c|c|c|c| } 
 \hline
  & ${\rm d}=1$ & ${\rm d}=2$ & ${\rm d}=3$ \\ 
 \hline
 $\alpha_{\rm L}$ & 0.5 & 0.18 & 0.09 \\ 
 $\alpha_{\rm NL}$ & 1 & 0.48 & 0.32\\
 $\alpha_{\rm KPZ}$ & 1/2 &0.39 & 0.31 \\
 \hline
\end{tabular}
\begin{tabular}{ |c|c|c|c| } 
 \hline
  & ${\rm d}=1$ & ${\rm d}=2$ & ${\rm d}=3$ \\ 
 \hline
 
  $\nu_{\rm L}$ & 0.3 & 0.21 & 0.16 \\ 
 $\nu_{\rm NL}$ & 1.41 & 1.46 & 1.83\\
 $\beta$ & 0.37 & 0.3 & 0.25 \\ 
 $\beta_{\rm KPZ}$ & 1/3 & 0.24 & 0.18 \\
 \hline
\end{tabular}
\begin{comment}
\begin{tabular}{ |c|c|c|c| } 
 \hline
  & ${\rm d}=1$ & ${\rm d}=2$ & ${\rm d}=3$ \\ 
 \hline
 $\nu_{\rm L}$ & 0.3 & 0.21 & 0.16 \\ 
 $\nu_{\rm NL}$ & 1.41 & 1.46 & 1.83\\
 \hline
\end{tabular}
\end{comment}
\end{center}
\caption{Different values of the roughness exponent for the local Eq.~\eqref{eq:localdiscreteequation} (L) and non-local Eq.~\eqref{eq:fundamentalequationpositionbasis} (NL) evolutions. For the real KPZ class, the values for $\alpha$ are taken from \citep{NumericsKPZRSOSsurface} and $\beta$ from the relation $\alpha + \frac{\alpha}{\beta} = 2$. Remark that $\alpha_{\rm L}$ and $\beta$ are consistent with the known coefficient of the KPZ class only for ${\rm d}=1$.} 
\label{table:roughness}
\end{table}

% \paragraph{Martin-Siggia-Rose-Janssen-De Dominicis (MSRJD) action }
\textit{Martin-Siggia-Rose-Janssen-De Dominicis (MSRJD) action }
We will now analyze our phase transition with field theory methods by deriving the MSRJD action \citep{MSRoriginalpaper,kamenev_2011,dominicis1976technics,janssen1976lagrangean}
$Z$ associated to (\ref{eq:localcontinuousequation}). The details
of the derivation are presented in the SM \citep{SM}. Let the superscript $a$ denote
the auxiliary fields. We have that 
\begin{equation}
Z=\int{\cal D}[\varphi,\bar{\varphi},\varphi^{a},\bar{\varphi}^{a}]e^{iS_{0}+iS_{\nu}},
\end{equation}
where the bar denotes complex conjugation, and $S_{0}$, $S_{\nu}$
are respectively the quadratic and quartic part of the action: 
\begin{comment}
(!!!Careful,
in the notes we use $z=\frac{\varphi}{\sqrt{2}}$!!!)
\end{comment}
\begin{align}
S_{0} & =\int d^{{\rm d}}\vec{r}dt(\bar{\varphi},\bar{\varphi}^{a})\boldsymbol{G}_{0}^{-1}\begin{pmatrix}\varphi\\
\varphi^{a}
\end{pmatrix},\\
\boldsymbol{G}_{0}^{-1} & =\frac{1}{2}\begin{pmatrix}0 & -\frac{\gamma}{2}+\partial_{t}-iD\nabla^{2}\\
-\frac{\gamma}{2}-\partial_{t}+iD\nabla^{2} & 0
\end{pmatrix},\label{eq:InverseG0}
\end{align}
\begin{equation}
S_{\nu}=\frac{i}{8}\int d^{{\rm d}}\vec{r}dt\big(\lambda^{{\rm I}}\left(\bar{\varphi}^{a}\right)^{2}\varphi^{2}+\lambda^{{\rm II}}\bar{\varphi}^{a}\bar{\varphi}\varphi^{a}\varphi+{\rm c.c}\big).
\label{eq:Snu}
\end{equation}
In \eqref{eq:Snu}, we introduced the labels I, II for the interacting terms, as they
will behave differently under renormalization. For the microscopic
theory \eqref{eq:localdiscreteequation}, we have $\lambda^{{\rm I}}=\lambda^{{\rm II}}=\lambda=\gamma b^{{\rm d}}$. 

Inverting (\ref{eq:InverseG0}) yields the free propagator in momentum
$\vec{q}$ and frequency $\omega$: 
\begin{align}
G_{0}^{R}(\vec{q},\omega) & =\frac{2i}{Dq^{2}-\omega-i\frac{\gamma}{2}}
\end{align}
where the $R$ label refers to the retarded propagator. The advanced
propagator $A$ is obtained by complex conjugation $G_{0}^{A}(\vec{q},\omega)=G_{0}^{R}(\vec{q},\omega)^*$.
\textit{Renormalization flow}
We proceed to the one-loop perturbative renormalization group (RG)
analysis of (\ref{eq:localcontinuousequation}). We employ standard
momentum-shell analysis \citep{kamenev_2011}. Let $\Lambda$ be the
microscopic momentum cut-off of the theory. The critical exponents
associated to $t$, $\varphi$, $\bar{\varphi}$ are named respectively
$z$, $\chi$ and $\bar{\chi}$. The flow is parametrized by $l$. 

At the one-loop level, there are no diagrams renormalizing the quadratic part
of the action.
Imposing the stationarity of the terms proportional to $\partial_{t}$ and $\nabla^{2}$ under the flow imposes
\begin{equation}
\chi+\bar{\chi}+{\rm d}=0,\quad z=2,
\end{equation}
and the renormalization of the "mass" term $\gamma$ is given by dimensional analysis: $\gamma=\gamma_0 e^{2l}$ where $\gamma_0$ is the bare value.
At one-loop level, the quartic interacting terms renormalize independently according to \citep{SM}: \begin{align}
%\frac{d\gamma}{dl}= & 2\gamma-{\rm sgn}\left(\gamma_{R}\right)K_{d}\left(2\lambda^{{\rm I}}+\lambda^{{\rm II}}\right),\nonumber \\
\frac{d\lambda^{{\rm I}}}{dl}= & (2-{\rm d})\lambda^{{\rm I}}+K_{\rm d}\frac{\left(\lambda^{\rm I}\right)^{2}}{\gamma+2i D\Lambda^{2}},\label{eq:flowequations}\\
\frac{d\lambda^{{\rm II}}}{dl}= & (2-{\rm d})\lambda^{{\rm II}}+K_{\rm d}\frac{\left(\lambda^{{\rm II}}\right)^{2}}{\gamma},\label{eq:flowlambdaII} 
\end{align}
%where ${\rm sgn}$ is the sign function, $\gamma_{R}:=\Re(\gamma)$
where we introduced $K_{{\rm d}}:=\frac{\Lambda^{{\rm d}}}{\Gamma({\rm d}/2)2^{{\rm d}-1}\pi^{{\rm d}/2}}$
with $\Gamma$ the Gamma function. Remark that even though the microscopic model is defined with real parameters, the flow takes $\lambda^{\rm I}$ to complex values. 

\begin{figure*}
\includegraphics[width=0.95
\textwidth]{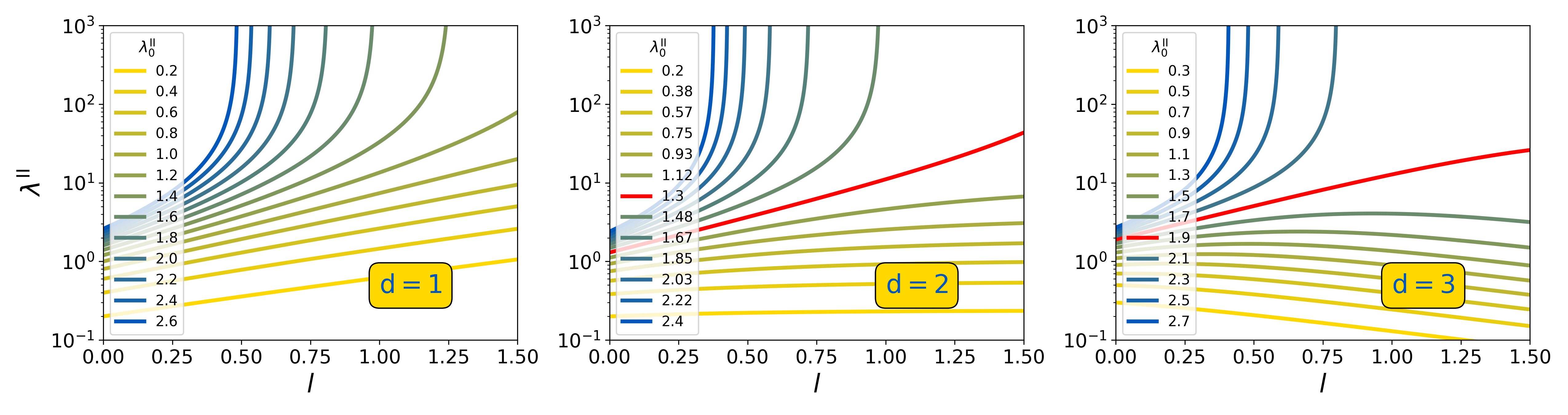}
\caption{Plots of Eq.~\eqref{eq:lambdaIIflowsolution} for different initial values $\lambda^{\rm II}_0$ and dimensions. For the plots, we fixed $a=1$, $\Lambda=\pi$, ${\gamma_0}=1$. The red curves in ${\rm d}=2$ and ${\rm d}=3$ indicate the critical values which separate the phase where $\lambda^{\rm II}\to \infty$ from the phase where it goes to either a finite (${\rm d}=2$) or $0$ value (${\rm d}=3$).}
\label{Fig:flowlambdII}
\end{figure*}

Starting from the same intial conditions for $\lambda^{\rm I}$ and $\lambda^{\rm II}$, we necessarily have $|\lambda^{{\rm I}}|\leq\lambda^{{\rm II}}$, hence we focus on $\lambda^{{\rm II}}$ to characterize the phase transition as it will be the first to diverge. Eq.~\eqref{eq:flowlambdaII} can be solved exactly: 

\begin{equation}
\lambda^{{\rm II}}=\lambda_{0}^{{\rm II}}\frac{e^{\left(2-{\rm d}\right)l}}{1-\frac{K_{{\rm d}}\lambda_{0}^{{\rm II}}}{{\rm d}\gamma_{0}}\left(1-e^{-{\rm d}l}\right)}.
\label{eq:lambdaIIflowsolution}
\end{equation}
For ${\rm d}=1$, the flow is always divergent as $l \to \infty$. For ${\rm d}>1$, the flow diverges for $\frac{K_{{\rm d}}\lambda_{0}^{{\rm II}}}{{\rm d}\gamma_{0}}\geq1$ while for $\frac{K_{{\rm d}}\lambda_{0}^{{\rm II}}}{{\rm d}\gamma_{0}}<1$ it either converges to a finite value (${\rm d}=2$) or to $0$ (${\rm d}\geq3$), see Fig.~\ref{Fig:flowlambdII}. We thus see that the lower critical dimension is $1$ (instead of $2$ for the real SHE equation), in agreement with the numerical computations. However, note that the microscopic value of $\frac{\lambda^{\rm II}_0}{\gamma_0}$ on which the behavior of the flow depends is \emph{fixed} in the simulations and can't be fine tuned accross the transition. More realistically, Eq.~\ref{eq:flowlambdaII} should contain a $D$ dependence that is not captured at one-loop level. Future investigations should aim at improving the RG scheme by, e.g., using dimensional regularization \citep{WieseSHE} or non-perturbative methods \citep{nonperturbativeRGLeonieCanet}.

% \paragraph{Conclusion }
\textit{Conclusion }
We have provided numerical and analytical arguments showing the existence
of a MiPT for a single free particle undergoing continuous measurement
when ${\rm d}>1$ and its absence when ${\rm d}=1$. Our work is one of the first to demonstrate the  critical role played by dimensionality to observe the existence of a transition or lack thereof in a quantum setting.
Compared to previous studies in the literature, it is remarkable that many-body effects play no role in the emergence of this transition. Additionally, for a single particle, the fermionic or bosonic statistics play no role. 

An important byproduct of our study is the local equation \eqref{eq:localcontinuousequation}, used as an approximation but interesting in itself. As mentioned above, it can be naturally interpreted as a complex version of the SHE equation. We showed however that its behavior departs significantly from its real counterpart as exemplified by the different lower critical dimension and the different critical exponents in the rough phase. This is particularly relevant in the light of the recent connection established between Anderson localization and a \emph{complex} directed polymer problem \citep{LemariePRLKPZAnderson}. This calls for a systematic study of complex SHE/directed polymer by allowing any value for the diffusion and noise parameter in the complex plane that we leave for future works.

Concerning future directions, one exciting possibility is that our transition is already visible
at the level of quantities linear in the density matrix, for instance
transport-related quantities. In 1d, it is known that measurements
induce a crossover from a ballistic to diffusive transport \citep{Znidaric__dephasingXXZ,Znidaric__XXdeph,Znidaric_dephasing,BauerBernardJinScipostfluctuatinghydro,BastianelloDeNardisDeluca_GHDDeph}
in free fermionic system. If we associate the ballistic behavior to
the delocalized phase and the diffusive behavior to the localized
one, a tempting conjecture is that this crossover  becomes a phase transition in higher dimensions. This route for characterizing MiPT would be particularly interesting for experiments, since the measurement of EE is computationally very heavy, often requiring tomography of the full quantum trajectories
and/or costly post-selection procedure \citep{NaturePhyMiPTExp,NatureMiPTexperiment}. 

Finally, on the theoretical side, we note that a recent interesting body of
literature has proposed non-linear sigma models as good effective
descriptions of free fermionic or spin chains under measurements \citep{NahumNLsM,LudwigNLsM,MirlinNLsM}. An alternative route proposed in these works to get the action is to write the stochastic Keldysh action associated to the tight-binding plus measurements model and to use replicas to study the fluctuations. It would be instructive to  understand whether this approach is compatible with our MSRJD action in the single-body limit.
 
\textbf{Acknowledgments}
Both authors express their gratitude towards Léonie Canet, Xhek Turkeshi and Gabriel Artur Weiderpass for
illuminating discussions. Part of this project was developed at ``\emph{Les
Gustins''} summer school. 

\textit{Note added}.---During the completion of this manuscript, two additional works studying MiPT in free fermions in 2 dimensions through the lenses of replica field theories and effective non-linear sigma models came to our awareness \citep{MirlinGornyiMiPT3DFreefermions,BuchholdMiPT2d}.
\bibliography{bibliosinglebodymeasures}

\begin{thebibliography}{53}%
\makeatletter
\providecommand \@ifxundefined [1]{%
 \@ifx{#1\undefined}
}%
\providecommand \@ifnum [1]{%
 \ifnum #1\expandafter \@firstoftwo
 \else \expandafter \@secondoftwo
 \fi
}%
\providecommand \@ifx [1]{%
 \ifx #1\expandafter \@firstoftwo
 \else \expandafter \@secondoftwo
 \fi
}%
\providecommand \natexlab [1]{#1}%
\providecommand \enquote  [1]{``#1''}%
\providecommand \bibnamefont  [1]{#1}%
\providecommand \bibfnamefont [1]{#1}%
\providecommand \citenamefont [1]{#1}%
\providecommand \href@noop [0]{\@secondoftwo}%
\providecommand \href [0]{\begingroup \@sanitize@url \@href}%
\providecommand \@href[1]{\@@startlink{#1}\@@href}%
\providecommand \@@href[1]{\endgroup#1\@@endlink}%
\providecommand \@sanitize@url [0]{\catcode `\\12\catcode `\$12\catcode
  `\&12\catcode `\#12\catcode `\^12\catcode `\_12\catcode `\%12\relax}%
\providecommand \@@startlink[1]{}%
\providecommand \@@endlink[0]{}%
\providecommand \url  [0]{\begingroup\@sanitize@url \@url }%
\providecommand \@url [1]{\endgroup\@href {#1}{\urlprefix }}%
\providecommand \urlprefix  [0]{URL }%
\providecommand \Eprint [0]{\href }%
\providecommand \doibase [0]{https://doi.org/}%
\providecommand \selectlanguage [0]{\@gobble}%
\providecommand \bibinfo  [0]{\@secondoftwo}%
\providecommand \bibfield  [0]{\@secondoftwo}%
\providecommand \translation [1]{[#1]}%
\providecommand \BibitemOpen [0]{}%
\providecommand \bibitemStop [0]{}%
\providecommand \bibitemNoStop [0]{.\EOS\space}%
\providecommand \EOS [0]{\spacefactor3000\relax}%
\providecommand \BibitemShut  [1]{\csname bibitem#1\endcsname}%
\let\auto@bib@innerbib\@empty
%</preamble>
\bibitem [{\citenamefont {Skinner}\ \emph {et~al.}(2019)\citenamefont
  {Skinner}, \citenamefont {Ruhman},\ and\ \citenamefont
  {Nahum}}]{NahumMeasurementinducedtransition2019}%
  \BibitemOpen
  \bibfield  {author} {\bibinfo {author} {\bibfnamefont {B.}~\bibnamefont
  {Skinner}}, \bibinfo {author} {\bibfnamefont {J.}~\bibnamefont {Ruhman}},\
  and\ \bibinfo {author} {\bibfnamefont {A.}~\bibnamefont {Nahum}},\ }\bibfield
   {title} {\bibinfo {title} {Measurement-{{Induced Phase Transitions}} in the
  {{Dynamics}} of {{Entanglement}}},\ }\href
  {https://doi.org/10.1103/PhysRevX.9.031009} {\bibfield  {journal} {\bibinfo
  {journal} {Phys. Rev. X}\ }\textbf {\bibinfo {volume} {9}},\ \bibinfo {pages}
  {031009} (\bibinfo {year} {2019})}\BibitemShut {NoStop}%
\bibitem [{\citenamefont {Szyniszewski}\ \emph {et~al.}(2019)\citenamefont
  {Szyniszewski}, \citenamefont {Romito},\ and\ \citenamefont
  {Schomerus}}]{MeasurementinducedSchomerus}%
  \BibitemOpen
  \bibfield  {author} {\bibinfo {author} {\bibfnamefont {M.}~\bibnamefont
  {Szyniszewski}}, \bibinfo {author} {\bibfnamefont {A.}~\bibnamefont
  {Romito}},\ and\ \bibinfo {author} {\bibfnamefont {H.}~\bibnamefont
  {Schomerus}},\ }\bibfield  {title} {\bibinfo {title} {Entanglement transition
  from variable-strength weak measurements},\ }\href
  {https://doi.org/10.1103/PhysRevB.100.064204} {\bibfield  {journal} {\bibinfo
   {journal} {Phys. Rev. B}\ }\textbf {\bibinfo {volume} {100}},\ \bibinfo
  {pages} {064204} (\bibinfo {year} {2019})}\BibitemShut {NoStop}%
\bibitem [{\citenamefont {Li}\ \emph {et~al.}(2019)\citenamefont {Li},
  \citenamefont {Chen},\ and\ \citenamefont
  {Fisher}}]{MeasurementinducedFisher}%
  \BibitemOpen
  \bibfield  {author} {\bibinfo {author} {\bibfnamefont {Y.}~\bibnamefont
  {Li}}, \bibinfo {author} {\bibfnamefont {X.}~\bibnamefont {Chen}},\ and\
  \bibinfo {author} {\bibfnamefont {M.~P.~A.}\ \bibnamefont {Fisher}},\
  }\bibfield  {title} {\bibinfo {title} {Measurement-driven entanglement
  transition in hybrid quantum circuits},\ }\href
  {https://doi.org/10.1103/PhysRevB.100.134306} {\bibfield  {journal} {\bibinfo
   {journal} {Phys. Rev. B}\ }\textbf {\bibinfo {volume} {100}},\ \bibinfo
  {pages} {134306} (\bibinfo {year} {2019})}\BibitemShut {NoStop}%
\bibitem [{\citenamefont {Turkeshi}\ \emph {et~al.}(2020)\citenamefont
  {Turkeshi}, \citenamefont {Fazio},\ and\ \citenamefont
  {Dalmonte}}]{TurkeShiMiPThybrid}%
  \BibitemOpen
  \bibfield  {author} {\bibinfo {author} {\bibfnamefont {X.}~\bibnamefont
  {Turkeshi}}, \bibinfo {author} {\bibfnamefont {R.}~\bibnamefont {Fazio}},\
  and\ \bibinfo {author} {\bibfnamefont {M.}~\bibnamefont {Dalmonte}},\
  }\bibfield  {title} {\bibinfo {title} {Measurement-induced criticality in
  $(2+1)$-dimensional hybrid quantum circuits},\ }\href
  {https://doi.org/10.1103/PhysRevB.102.014315} {\bibfield  {journal} {\bibinfo
   {journal} {Phys. Rev. B}\ }\textbf {\bibinfo {volume} {102}},\ \bibinfo
  {pages} {014315} (\bibinfo {year} {2020})}\BibitemShut {NoStop}%
\bibitem [{\citenamefont {Gullans}\ and\ \citenamefont
  {Huse}(2020{\natexlab{a}})}]{HuseMeasurementinduced}%
  \BibitemOpen
  \bibfield  {author} {\bibinfo {author} {\bibfnamefont {M.~J.}\ \bibnamefont
  {Gullans}}\ and\ \bibinfo {author} {\bibfnamefont {D.~A.}\ \bibnamefont
  {Huse}},\ }\bibfield  {title} {\bibinfo {title} {Scalable probes of
  measurement-induced criticality},\ }\href
  {https://doi.org/10.1103/PhysRevLett.125.070606} {\bibfield  {journal}
  {\bibinfo  {journal} {Phys. Rev. Lett.}\ }\textbf {\bibinfo {volume} {125}},\
  \bibinfo {pages} {070606} (\bibinfo {year} {2020}{\natexlab{a}})}\BibitemShut
  {NoStop}%
\bibitem [{\citenamefont {Gullans}\ and\ \citenamefont
  {Huse}(2020{\natexlab{b}})}]{HuseMeasurementinduced3}%
  \BibitemOpen
  \bibfield  {author} {\bibinfo {author} {\bibfnamefont {M.~J.}\ \bibnamefont
  {Gullans}}\ and\ \bibinfo {author} {\bibfnamefont {D.~A.}\ \bibnamefont
  {Huse}},\ }\bibfield  {title} {\bibinfo {title} {Dynamical purification phase
  transition induced by quantum measurements},\ }\href
  {https://doi.org/10.1103/PhysRevX.10.041020} {\bibfield  {journal} {\bibinfo
  {journal} {Phys. Rev. X}\ }\textbf {\bibinfo {volume} {10}},\ \bibinfo
  {pages} {041020} (\bibinfo {year} {2020}{\natexlab{b}})}\BibitemShut
  {NoStop}%
\bibitem [{\citenamefont {Zabalo}\ \emph {et~al.}(2020)\citenamefont {Zabalo},
  \citenamefont {Gullans}, \citenamefont {Wilson}, \citenamefont
  {Gopalakrishnan}, \citenamefont {Huse},\ and\ \citenamefont
  {Pixley}}]{HuseMeasurementinduced2}%
  \BibitemOpen
  \bibfield  {author} {\bibinfo {author} {\bibfnamefont {A.}~\bibnamefont
  {Zabalo}}, \bibinfo {author} {\bibfnamefont {M.~J.}\ \bibnamefont {Gullans}},
  \bibinfo {author} {\bibfnamefont {J.~H.}\ \bibnamefont {Wilson}}, \bibinfo
  {author} {\bibfnamefont {S.}~\bibnamefont {Gopalakrishnan}}, \bibinfo
  {author} {\bibfnamefont {D.~A.}\ \bibnamefont {Huse}},\ and\ \bibinfo
  {author} {\bibfnamefont {J.~H.}\ \bibnamefont {Pixley}},\ }\bibfield  {title}
  {\bibinfo {title} {Critical properties of the measurement-induced transition
  in random quantum circuits},\ }\href
  {https://doi.org/10.1103/PhysRevB.101.060301} {\bibfield  {journal} {\bibinfo
   {journal} {Phys. Rev. B}\ }\textbf {\bibinfo {volume} {101}},\ \bibinfo
  {pages} {060301} (\bibinfo {year} {2020})}\BibitemShut {NoStop}%
\bibitem [{\citenamefont {Bao}\ \emph {et~al.}(2020)\citenamefont {Bao},
  \citenamefont {Choi},\ and\ \citenamefont
  {Altman}}]{MeasurementinducedAltman}%
  \BibitemOpen
  \bibfield  {author} {\bibinfo {author} {\bibfnamefont {Y.}~\bibnamefont
  {Bao}}, \bibinfo {author} {\bibfnamefont {S.}~\bibnamefont {Choi}},\ and\
  \bibinfo {author} {\bibfnamefont {E.}~\bibnamefont {Altman}},\ }\bibfield
  {title} {\bibinfo {title} {Theory of the phase transition in random unitary
  circuits with measurements},\ }\href
  {https://doi.org/10.1103/PhysRevB.101.104301} {\bibfield  {journal} {\bibinfo
   {journal} {Phys. Rev. B}\ }\textbf {\bibinfo {volume} {101}},\ \bibinfo
  {pages} {104301} (\bibinfo {year} {2020})}\BibitemShut {NoStop}%
\bibitem [{\citenamefont {Turkeshi}\ \emph {et~al.}(2021)\citenamefont
  {Turkeshi}, \citenamefont {Biella}, \citenamefont {Fazio}, \citenamefont
  {Dalmonte},\ and\ \citenamefont {Schir\'o}}]{TurkeshiMiPTinfinitezeroclick}%
  \BibitemOpen
  \bibfield  {author} {\bibinfo {author} {\bibfnamefont {X.}~\bibnamefont
  {Turkeshi}}, \bibinfo {author} {\bibfnamefont {A.}~\bibnamefont {Biella}},
  \bibinfo {author} {\bibfnamefont {R.}~\bibnamefont {Fazio}}, \bibinfo
  {author} {\bibfnamefont {M.}~\bibnamefont {Dalmonte}},\ and\ \bibinfo
  {author} {\bibfnamefont {M.}~\bibnamefont {Schir\'o}},\ }\bibfield  {title}
  {\bibinfo {title} {Measurement-induced entanglement transitions in the
  quantum ising chain: From infinite to zero clicks},\ }\href
  {https://doi.org/10.1103/PhysRevB.103.224210} {\bibfield  {journal} {\bibinfo
   {journal} {Phys. Rev. B}\ }\textbf {\bibinfo {volume} {103}},\ \bibinfo
  {pages} {224210} (\bibinfo {year} {2021})}\BibitemShut {NoStop}%
\bibitem [{\citenamefont {Weinstein}\ \emph {et~al.}(2022)\citenamefont
  {Weinstein}, \citenamefont {Bao},\ and\ \citenamefont
  {Altman}}]{weinstein2022measurement}%
  \BibitemOpen
  \bibfield  {author} {\bibinfo {author} {\bibfnamefont {Z.}~\bibnamefont
  {Weinstein}}, \bibinfo {author} {\bibfnamefont {Y.}~\bibnamefont {Bao}},\
  and\ \bibinfo {author} {\bibfnamefont {E.}~\bibnamefont {Altman}},\
  }\bibfield  {title} {\bibinfo {title} {Measurement-induced power-law
  negativity in an open monitored quantum circuit},\ }\href@noop {} {\bibfield
  {journal} {\bibinfo  {journal} {Physical Review Letters}\ }\textbf {\bibinfo
  {volume} {129}},\ \bibinfo {pages} {080501} (\bibinfo {year}
  {2022})}\BibitemShut {NoStop}%
\bibitem [{\citenamefont {Minoguchi}\ \emph {et~al.}(2022)\citenamefont
  {Minoguchi}, \citenamefont {Rabl},\ and\ \citenamefont
  {Buchhold}}]{MeasurementBosonsRabl}%
  \BibitemOpen
  \bibfield  {author} {\bibinfo {author} {\bibfnamefont {Y.}~\bibnamefont
  {Minoguchi}}, \bibinfo {author} {\bibfnamefont {P.}~\bibnamefont {Rabl}},\
  and\ \bibinfo {author} {\bibfnamefont {M.}~\bibnamefont {Buchhold}},\
  }\bibfield  {title} {\bibinfo {title} {{Continuous gaussian measurements of
  the free boson CFT: A model for exactly solvable and detectable
  measurement-induced dynamics}},\ }\href
  {https://doi.org/10.21468/SciPostPhys.12.1.009} {\bibfield  {journal}
  {\bibinfo  {journal} {SciPost Phys.}\ }\textbf {\bibinfo {volume} {12}},\
  \bibinfo {pages} {009} (\bibinfo {year} {2022})}\BibitemShut {NoStop}%
\bibitem [{\citenamefont {Gal}\ \emph {et~al.}(2023)\citenamefont {Gal},
  \citenamefont {Turkeshi},\ and\ \citenamefont
  {Schirò}}]{YouennXhekSchiroMIPTnonHermitian}%
  \BibitemOpen
  \bibfield  {author} {\bibinfo {author} {\bibfnamefont {Y.~L.}\ \bibnamefont
  {Gal}}, \bibinfo {author} {\bibfnamefont {X.}~\bibnamefont {Turkeshi}},\ and\
  \bibinfo {author} {\bibfnamefont {M.}~\bibnamefont {Schirò}},\ }\bibfield
  {title} {\bibinfo {title} {{Volume-to-area law entanglement transition in a
  non-Hermitian free fermionic chain}},\ }\href
  {https://doi.org/10.21468/SciPostPhys.14.5.138} {\bibfield  {journal}
  {\bibinfo  {journal} {SciPost Phys.}\ }\textbf {\bibinfo {volume} {14}},\
  \bibinfo {pages} {138} (\bibinfo {year} {2023})}\BibitemShut {NoStop}%
\bibitem [{\citenamefont {Granet}\ \emph {et~al.}(2023)\citenamefont {Granet},
  \citenamefont {Zhang},\ and\ \citenamefont
  {Dreyer}}]{GranetMiPTPostSelection}%
  \BibitemOpen
  \bibfield  {author} {\bibinfo {author} {\bibfnamefont {E.}~\bibnamefont
  {Granet}}, \bibinfo {author} {\bibfnamefont {C.}~\bibnamefont {Zhang}},\ and\
  \bibinfo {author} {\bibfnamefont {H.}~\bibnamefont {Dreyer}},\ }\bibfield
  {title} {\bibinfo {title} {Volume-law to area-law entanglement transition in
  a nonunitary periodic gaussian circuit},\ }\href
  {https://doi.org/10.1103/PhysRevLett.130.230401} {\bibfield  {journal}
  {\bibinfo  {journal} {Phys. Rev. Lett.}\ }\textbf {\bibinfo {volume} {130}},\
  \bibinfo {pages} {230401} (\bibinfo {year} {2023})}\BibitemShut {NoStop}%
\bibitem [{\citenamefont {{Lee}}\ \emph {et~al.}(2023)\citenamefont {{Lee}},
  \citenamefont {{Jin}}, \citenamefont {{Wang}}, \citenamefont {{McDonald}},\
  and\ \citenamefont {{Clerk}}}]{GideonEEBKC}%
  \BibitemOpen
  \bibfield  {author} {\bibinfo {author} {\bibfnamefont {G.}~\bibnamefont
  {{Lee}}}, \bibinfo {author} {\bibfnamefont {T.}~\bibnamefont {{Jin}}},
  \bibinfo {author} {\bibfnamefont {Y.-X.}\ \bibnamefont {{Wang}}}, \bibinfo
  {author} {\bibfnamefont {A.}~\bibnamefont {{McDonald}}},\ and\ \bibinfo
  {author} {\bibfnamefont {A.}~\bibnamefont {{Clerk}}},\ }\bibfield  {title}
  {\bibinfo {title} {{Entanglement phase transition due to reciprocity breaking
  without measurement or post-selection}},\ }\href
  {https://doi.org/10.48550/arXiv.2308.14614} {\bibfield  {journal} {\bibinfo
  {journal} {arXiv e-prints}\ ,\ \bibinfo {eid} {arXiv:2308.14614}} (\bibinfo
  {year} {2023})},\ \Eprint {https://arxiv.org/abs/2308.14614}
  {arXiv:2308.14614 [quant-ph]} \BibitemShut {NoStop}%
\bibitem [{\citenamefont {Cao}\ \emph {et~al.}(2019)\citenamefont {Cao},
  \citenamefont {Tilloy},\ and\ \citenamefont
  {De~Luca}}]{caoEntanglementFermionChain2019}%
  \BibitemOpen
  \bibfield  {author} {\bibinfo {author} {\bibfnamefont {X.}~\bibnamefont
  {Cao}}, \bibinfo {author} {\bibfnamefont {A.}~\bibnamefont {Tilloy}},\ and\
  \bibinfo {author} {\bibfnamefont {A.}~\bibnamefont {De~Luca}},\ }\bibfield
  {title} {\bibinfo {title} {Entanglement in a fermion chain under continuous
  monitoring},\ }\href {https://doi.org/10.21468/SciPostPhys.7.2.024}
  {\bibfield  {journal} {\bibinfo  {journal} {SciPost Physics}\ }\textbf
  {\bibinfo {volume} {7}},\ \bibinfo {pages} {024} (\bibinfo {year}
  {2019})}\BibitemShut {NoStop}%
\bibitem [{\citenamefont {Alberton}\ \emph {et~al.}(2021)\citenamefont
  {Alberton}, \citenamefont {Buchhold},\ and\ \citenamefont
  {Diehl}}]{BuchholdMiPTPRL}%
  \BibitemOpen
  \bibfield  {author} {\bibinfo {author} {\bibfnamefont {O.}~\bibnamefont
  {Alberton}}, \bibinfo {author} {\bibfnamefont {M.}~\bibnamefont {Buchhold}},\
  and\ \bibinfo {author} {\bibfnamefont {S.}~\bibnamefont {Diehl}},\ }\bibfield
   {title} {\bibinfo {title} {Entanglement transition in a monitored
  free-fermion chain: From extended criticality to area law},\ }\href
  {https://doi.org/10.1103/PhysRevLett.126.170602} {\bibfield  {journal}
  {\bibinfo  {journal} {Phys. Rev. Lett.}\ }\textbf {\bibinfo {volume} {126}},\
  \bibinfo {pages} {170602} (\bibinfo {year} {2021})}\BibitemShut {NoStop}%
\bibitem [{\citenamefont {Buchhold}\ \emph {et~al.}(2021)\citenamefont
  {Buchhold}, \citenamefont {Minoguchi}, \citenamefont {Altland},\ and\
  \citenamefont {Diehl}}]{BuchholdMiPTPRX}%
  \BibitemOpen
  \bibfield  {author} {\bibinfo {author} {\bibfnamefont {M.}~\bibnamefont
  {Buchhold}}, \bibinfo {author} {\bibfnamefont {Y.}~\bibnamefont {Minoguchi}},
  \bibinfo {author} {\bibfnamefont {A.}~\bibnamefont {Altland}},\ and\ \bibinfo
  {author} {\bibfnamefont {S.}~\bibnamefont {Diehl}},\ }\bibfield  {title}
  {\bibinfo {title} {Effective theory for the measurement-induced phase
  transition of dirac fermions},\ }\href
  {https://doi.org/10.1103/PhysRevX.11.041004} {\bibfield  {journal} {\bibinfo
  {journal} {Phys. Rev. X}\ }\textbf {\bibinfo {volume} {11}},\ \bibinfo
  {pages} {041004} (\bibinfo {year} {2021})}\BibitemShut {NoStop}%
\bibitem [{\citenamefont {{Jian}}\ \emph {et~al.}(2023)\citenamefont {{Jian}},
  \citenamefont {{Shapourian}}, \citenamefont {{Bauer}},\ and\ \citenamefont
  {{Ludwig}}}]{LudwigNLsM}%
  \BibitemOpen
  \bibfield  {author} {\bibinfo {author} {\bibfnamefont {C.-M.}\ \bibnamefont
  {{Jian}}}, \bibinfo {author} {\bibfnamefont {H.}~\bibnamefont
  {{Shapourian}}}, \bibinfo {author} {\bibfnamefont {B.}~\bibnamefont
  {{Bauer}}},\ and\ \bibinfo {author} {\bibfnamefont {A.~W.~W.}\ \bibnamefont
  {{Ludwig}}},\ }\bibfield  {title} {\bibinfo {title} {{Measurement-induced
  entanglement transitions in quantum circuits of non-interacting fermions:
  Born-rule versus forced measurements}},\ }\href
  {https://doi.org/10.48550/arXiv.2302.09094} {\bibfield  {journal} {\bibinfo
  {journal} {arXiv e-prints}\ ,\ \bibinfo {eid} {arXiv:2302.09094}} (\bibinfo
  {year} {2023})},\ \Eprint {https://arxiv.org/abs/2302.09094}
  {arXiv:2302.09094 [cond-mat.stat-mech]} \BibitemShut {NoStop}%
\bibitem [{\citenamefont {{Fava}}\ \emph {et~al.}(2023)\citenamefont {{Fava}},
  \citenamefont {{Piroli}}, \citenamefont {{Swann}}, \citenamefont
  {{Bernard}},\ and\ \citenamefont {{Nahum}}}]{NahumNLsM}%
  \BibitemOpen
  \bibfield  {author} {\bibinfo {author} {\bibfnamefont {M.}~\bibnamefont
  {{Fava}}}, \bibinfo {author} {\bibfnamefont {L.}~\bibnamefont {{Piroli}}},
  \bibinfo {author} {\bibfnamefont {T.}~\bibnamefont {{Swann}}}, \bibinfo
  {author} {\bibfnamefont {D.}~\bibnamefont {{Bernard}}},\ and\ \bibinfo
  {author} {\bibfnamefont {A.}~\bibnamefont {{Nahum}}},\ }\bibfield  {title}
  {\bibinfo {title} {{Nonlinear sigma models for monitored dynamics of free
  fermions}},\ }\href {https://doi.org/10.48550/arXiv.2302.12820} {\bibfield
  {journal} {\bibinfo  {journal} {arXiv e-prints}\ ,\ \bibinfo {eid}
  {arXiv:2302.12820}} (\bibinfo {year} {2023})},\ \Eprint
  {https://arxiv.org/abs/2302.12820} {arXiv:2302.12820 [cond-mat.stat-mech]}
  \BibitemShut {NoStop}%
\bibitem [{\citenamefont {{Poboiko}}\ \emph
  {et~al.}(2023{\natexlab{a}})\citenamefont {{Poboiko}}, \citenamefont
  {{P{\"o}pperl}}, \citenamefont {{Gornyi}},\ and\ \citenamefont
  {{Mirlin}}}]{MirlinNLsM}%
  \BibitemOpen
  \bibfield  {author} {\bibinfo {author} {\bibfnamefont {I.}~\bibnamefont
  {{Poboiko}}}, \bibinfo {author} {\bibfnamefont {P.}~\bibnamefont
  {{P{\"o}pperl}}}, \bibinfo {author} {\bibfnamefont {I.~V.}\ \bibnamefont
  {{Gornyi}}},\ and\ \bibinfo {author} {\bibfnamefont {A.~D.}\ \bibnamefont
  {{Mirlin}}},\ }\bibfield  {title} {\bibinfo {title} {{Theory of free fermions
  under random projective measurements}},\ }\href
  {https://doi.org/10.48550/arXiv.2304.03138} {\bibfield  {journal} {\bibinfo
  {journal} {arXiv e-prints}\ ,\ \bibinfo {eid} {arXiv:2304.03138}} (\bibinfo
  {year} {2023}{\natexlab{a}})},\ \Eprint {https://arxiv.org/abs/2304.03138}
  {arXiv:2304.03138 [quant-ph]} \BibitemShut {NoStop}%
\bibitem [{\citenamefont {L\'oio}\ \emph {et~al.}(2023)\citenamefont {L\'oio},
  \citenamefont {De~Luca}, \citenamefont {De~Nardis},\ and\ \citenamefont
  {Turkeshi}}]{PRBPurificationtimescale}%
  \BibitemOpen
  \bibfield  {author} {\bibinfo {author} {\bibfnamefont {H.}~\bibnamefont
  {L\'oio}}, \bibinfo {author} {\bibfnamefont {A.}~\bibnamefont {De~Luca}},
  \bibinfo {author} {\bibfnamefont {J.}~\bibnamefont {De~Nardis}},\ and\
  \bibinfo {author} {\bibfnamefont {X.}~\bibnamefont {Turkeshi}},\ }\bibfield
  {title} {\bibinfo {title} {Purification timescales in monitored fermions},\
  }\href {https://doi.org/10.1103/PhysRevB.108.L020306} {\bibfield  {journal}
  {\bibinfo  {journal} {Phys. Rev. B}\ }\textbf {\bibinfo {volume} {108}},\
  \bibinfo {pages} {L020306} (\bibinfo {year} {2023})}\BibitemShut {NoStop}%
\bibitem [{\citenamefont {Jin}\ and\ \citenamefont
  {Martin}(2022)}]{JinMartinMiPTclassical}%
  \BibitemOpen
  \bibfield  {author} {\bibinfo {author} {\bibfnamefont {T.}~\bibnamefont
  {Jin}}\ and\ \bibinfo {author} {\bibfnamefont {D.~G.}\ \bibnamefont
  {Martin}},\ }\bibfield  {title} {\bibinfo {title} {Kardar-parisi-zhang
  physics and phase transition in a classical single random walker under
  continuous measurement},\ }\href
  {https://doi.org/10.1103/PhysRevLett.129.260603} {\bibfield  {journal}
  {\bibinfo  {journal} {Phys. Rev. Lett.}\ }\textbf {\bibinfo {volume} {129}},\
  \bibinfo {pages} {260603} (\bibinfo {year} {2022})}\BibitemShut {NoStop}%
\bibitem [{\citenamefont {Jacobs}\ and\ \citenamefont
  {Steck}(2006)}]{JacobsintroductiontoCmeasurement}%
  \BibitemOpen
  \bibfield  {author} {\bibinfo {author} {\bibfnamefont {K.}~\bibnamefont
  {Jacobs}}\ and\ \bibinfo {author} {\bibfnamefont {D.~A.}\ \bibnamefont
  {Steck}},\ }\bibfield  {title} {\bibinfo {title} {A straightforward
  introduction to continuous quantum measurement},\ }\href
  {https://doi.org/10.1080/00107510601101934} {\bibfield  {journal} {\bibinfo
  {journal} {Contemporary Physics}\ }\textbf {\bibinfo {volume} {47}},\
  \bibinfo {pages} {279} (\bibinfo {year} {2006})},\ \Eprint
  {https://arxiv.org/abs/https://doi.org/10.1080/00107510601101934}
  {https://doi.org/10.1080/00107510601101934} \BibitemShut {NoStop}%
\bibitem [{\citenamefont {Bernard}\ \emph {et~al.}(2018)\citenamefont
  {Bernard}, \citenamefont {Jin},\ and\ \citenamefont
  {Shpielberg}}]{BernardJinShpielberg_2018}%
  \BibitemOpen
  \bibfield  {author} {\bibinfo {author} {\bibfnamefont {D.}~\bibnamefont
  {Bernard}}, \bibinfo {author} {\bibfnamefont {T.}~\bibnamefont {Jin}},\ and\
  \bibinfo {author} {\bibfnamefont {O.}~\bibnamefont {Shpielberg}},\ }\bibfield
   {title} {\bibinfo {title} {Transport in quantum chains under strong
  monitoring},\ }\href {https://doi.org/10.1209/0295-5075/121/60006} {\bibfield
   {journal} {\bibinfo  {journal} {{EPL} (Europhysics Letters)}\ }\textbf
  {\bibinfo {volume} {121}},\ \bibinfo {pages} {60006} (\bibinfo {year}
  {2018})}\BibitemShut {NoStop}%
\bibitem [{\citenamefont {Carisch}\ \emph {et~al.}(2023)\citenamefont
  {Carisch}, \citenamefont {Romito},\ and\ \citenamefont
  {Zilberberg}}]{Oded1Dmeasurement}%
  \BibitemOpen
  \bibfield  {author} {\bibinfo {author} {\bibfnamefont {C.}~\bibnamefont
  {Carisch}}, \bibinfo {author} {\bibfnamefont {A.}~\bibnamefont {Romito}},\
  and\ \bibinfo {author} {\bibfnamefont {O.}~\bibnamefont {Zilberberg}},\
  }\bibfield  {title} {\bibinfo {title} {Quantifying measurement-induced
  quantum-to-classical crossover using an open-system entanglement measure},\
  }\href@noop {} {\bibfield  {journal} {\bibinfo  {journal} {arXiv preprint
  arXiv:2304.02965}\ } (\bibinfo {year} {2023})}\BibitemShut {NoStop}%
\bibitem [{\citenamefont {Jin}\ \emph {et~al.}(2022)\citenamefont {Jin},
  \citenamefont {Ferreira}, \citenamefont {Filippone},\ and\ \citenamefont
  {Giamarchi}}]{JinTransportMeasures}%
  \BibitemOpen
  \bibfield  {author} {\bibinfo {author} {\bibfnamefont {T.}~\bibnamefont
  {Jin}}, \bibinfo {author} {\bibfnamefont {J.~a.~S.}\ \bibnamefont
  {Ferreira}}, \bibinfo {author} {\bibfnamefont {M.}~\bibnamefont
  {Filippone}},\ and\ \bibinfo {author} {\bibfnamefont {T.}~\bibnamefont
  {Giamarchi}},\ }\bibfield  {title} {\bibinfo {title} {Exact description of
  quantum stochastic models as quantum resistors},\ }\href
  {https://doi.org/10.1103/PhysRevResearch.4.013109} {\bibfield  {journal}
  {\bibinfo  {journal} {Phys. Rev. Res.}\ }\textbf {\bibinfo {volume} {4}},\
  \bibinfo {pages} {013109} (\bibinfo {year} {2022})}\BibitemShut {NoStop}%
\bibitem [{\citenamefont {{Ferreira}}\ \emph {et~al.}(2023)\citenamefont
  {{Ferreira}}, \citenamefont {{Jin}}, \citenamefont {{Mannhart}},
  \citenamefont {{Giamarchi}},\ and\ \citenamefont
  {{Filippone}}}]{FerreiraThermalenginesMeasure}%
  \BibitemOpen
  \bibfield  {author} {\bibinfo {author} {\bibfnamefont {J.}~\bibnamefont
  {{Ferreira}}}, \bibinfo {author} {\bibfnamefont {T.}~\bibnamefont {{Jin}}},
  \bibinfo {author} {\bibfnamefont {J.}~\bibnamefont {{Mannhart}}}, \bibinfo
  {author} {\bibfnamefont {T.}~\bibnamefont {{Giamarchi}}},\ and\ \bibinfo
  {author} {\bibfnamefont {M.}~\bibnamefont {{Filippone}}},\ }\bibfield
  {title} {\bibinfo {title} {{Exact description of transport and
  non-reciprocity in monitored quantum devices}},\ }\href
  {https://doi.org/10.48550/arXiv.2306.16452} {\bibfield  {journal} {\bibinfo
  {journal} {arXiv e-prints}\ ,\ \bibinfo {eid} {arXiv:2306.16452}} (\bibinfo
  {year} {2023})},\ \Eprint {https://arxiv.org/abs/2306.16452}
  {arXiv:2306.16452 [quant-ph]} \BibitemShut {NoStop}%
\bibitem [{\citenamefont {Kardar}\ \emph {et~al.}(1986)\citenamefont {Kardar},
  \citenamefont {Parisi},\ and\ \citenamefont {Zhang}}]{KPZpaper}%
  \BibitemOpen
  \bibfield  {author} {\bibinfo {author} {\bibfnamefont {M.}~\bibnamefont
  {Kardar}}, \bibinfo {author} {\bibfnamefont {G.}~\bibnamefont {Parisi}},\
  and\ \bibinfo {author} {\bibfnamefont {Y.-C.}\ \bibnamefont {Zhang}},\
  }\bibfield  {title} {\bibinfo {title} {Dynamic scaling of growing
  interfaces},\ }\href {https://doi.org/10.1103/PhysRevLett.56.889} {\bibfield
  {journal} {\bibinfo  {journal} {Phys. Rev. Lett.}\ }\textbf {\bibinfo
  {volume} {56}},\ \bibinfo {pages} {889} (\bibinfo {year} {1986})}\BibitemShut
  {NoStop}%
\bibitem [{\citenamefont {Bertini}\ and\ \citenamefont
  {Cancrini}(1995)}]{BertiniSHEFeynmanKac}%
  \BibitemOpen
  \bibfield  {author} {\bibinfo {author} {\bibfnamefont {L.}~\bibnamefont
  {Bertini}}\ and\ \bibinfo {author} {\bibfnamefont {N.}~\bibnamefont
  {Cancrini}},\ }\bibfield  {title} {\bibinfo {title} {The stochastic heat
  equation: Feynman-kac formula and intermittence},\ }\href
  {https://doi.org/10.1007/BF02180136} {\bibfield  {journal} {\bibinfo
  {journal} {Journal of Statistical Physics}\ }\textbf {\bibinfo {volume}
  {78}},\ \bibinfo {pages} {1377} (\bibinfo {year} {1995})}\BibitemShut
  {NoStop}%
\bibitem [{\citenamefont {Barab\'{a}si}\ and\ \citenamefont
  {Stanley}(1995)}]{FractalconceptsBarbarasi}%
  \BibitemOpen
  \bibfield  {author} {\bibinfo {author} {\bibfnamefont {A.-L.}\ \bibnamefont
  {Barab\'{a}si}}\ and\ \bibinfo {author} {\bibfnamefont {H.~E.}\ \bibnamefont
  {Stanley}},\ }\href {https://doi.org/10.1017/CBO9780511599798} {\emph
  {\bibinfo {title} {Fractal Concepts in Surface Growth}}}\ (\bibinfo
  {publisher} {Cambridge University Press},\ \bibinfo {year}
  {1995})\BibitemShut {NoStop}%
\bibitem [{\citenamefont {Family}\ and\ \citenamefont
  {Vicsek}(1985)}]{FamilyVicsek}%
  \BibitemOpen
  \bibfield  {author} {\bibinfo {author} {\bibfnamefont {F.}~\bibnamefont
  {Family}}\ and\ \bibinfo {author} {\bibfnamefont {T.}~\bibnamefont
  {Vicsek}},\ }\bibfield  {title} {\bibinfo {title} {Scaling of the active zone
  in the eden process on percolation networks and the ballistic deposition
  model},\ }\href {https://doi.org/10.1088/0305-4470/18/2/005} {\bibfield
  {journal} {\bibinfo  {journal} {Journal of Physics A: Mathematical and
  General}\ }\textbf {\bibinfo {volume} {18}},\ \bibinfo {pages} {L75}
  (\bibinfo {year} {1985})}\BibitemShut {NoStop}%
\bibitem [{\citenamefont {Forrest}\ and\ \citenamefont
  {Toral}(1993)}]{CrossoverKPZForrest}%
  \BibitemOpen
  \bibfield  {author} {\bibinfo {author} {\bibfnamefont {B.~M.}\ \bibnamefont
  {Forrest}}\ and\ \bibinfo {author} {\bibfnamefont {R.}~\bibnamefont
  {Toral}},\ }\bibfield  {title} {\bibinfo {title} {Crossover and finite-size
  effects in the (1+1)-dimensional kardar-parisi-zhang equation},\ }\href
  {https://doi.org/10.1007/BF01053591} {\bibfield  {journal} {\bibinfo
  {journal} {Journal of Statistical Physics}\ }\textbf {\bibinfo {volume}
  {70}},\ \bibinfo {pages} {703} (\bibinfo {year} {1993})}\BibitemShut
  {NoStop}%
\bibitem [{\citenamefont {Chame}\ and\ \citenamefont {Aar\~ao
  Reis}(2002)}]{CrossoverKPZChame}%
  \BibitemOpen
  \bibfield  {author} {\bibinfo {author} {\bibfnamefont {A.}~\bibnamefont
  {Chame}}\ and\ \bibinfo {author} {\bibfnamefont {F.~D.~A.}\ \bibnamefont
  {Aar\~ao Reis}},\ }\bibfield  {title} {\bibinfo {title} {Crossover effects in
  a discrete deposition model with kardar-parisi-zhang scaling},\ }\href
  {https://doi.org/10.1103/PhysRevE.66.051104} {\bibfield  {journal} {\bibinfo
  {journal} {Phys. Rev. E}\ }\textbf {\bibinfo {volume} {66}},\ \bibinfo
  {pages} {051104} (\bibinfo {year} {2002})}\BibitemShut {NoStop}%
\bibitem [{\citenamefont {Marinari}\ \emph {et~al.}(2000)\citenamefont
  {Marinari}, \citenamefont {Pagnani},\ and\ \citenamefont
  {Parisi}}]{NumericsKPZRSOSsurface}%
  \BibitemOpen
  \bibfield  {author} {\bibinfo {author} {\bibfnamefont {E.}~\bibnamefont
  {Marinari}}, \bibinfo {author} {\bibfnamefont {A.}~\bibnamefont {Pagnani}},\
  and\ \bibinfo {author} {\bibfnamefont {G.}~\bibnamefont {Parisi}},\
  }\bibfield  {title} {\bibinfo {title} {Critical exponents of the kpz equation
  via multi-surface coding numerical simulations},\ }\href
  {https://doi.org/10.1088/0305-4470/33/46/303} {\bibfield  {journal} {\bibinfo
   {journal} {Journal of Physics A: Mathematical and General}\ }\textbf
  {\bibinfo {volume} {33}},\ \bibinfo {pages} {8181} (\bibinfo {year}
  {2000})}\BibitemShut {NoStop}%
\bibitem [{\citenamefont {Martin}\ \emph {et~al.}(1973)\citenamefont {Martin},
  \citenamefont {Siggia},\ and\ \citenamefont {Rose}}]{MSRoriginalpaper}%
  \BibitemOpen
  \bibfield  {author} {\bibinfo {author} {\bibfnamefont {P.~C.}\ \bibnamefont
  {Martin}}, \bibinfo {author} {\bibfnamefont {E.~D.}\ \bibnamefont {Siggia}},\
  and\ \bibinfo {author} {\bibfnamefont {H.~A.}\ \bibnamefont {Rose}},\
  }\bibfield  {title} {\bibinfo {title} {Statistical dynamics of classical
  systems},\ }\href {https://doi.org/10.1103/PhysRevA.8.423} {\bibfield
  {journal} {\bibinfo  {journal} {Phys. Rev. A}\ }\textbf {\bibinfo {volume}
  {8}},\ \bibinfo {pages} {423} (\bibinfo {year} {1973})}\BibitemShut {NoStop}%
\bibitem [{\citenamefont {Kamenev}(2011)}]{kamenev_2011}%
  \BibitemOpen
  \bibfield  {author} {\bibinfo {author} {\bibfnamefont {A.}~\bibnamefont
  {Kamenev}},\ }\href {https://doi.org/10.1017/CBO9781139003667} {\emph
  {\bibinfo {title} {Field Theory of Non-Equilibrium Systems}}}\ (\bibinfo
  {publisher} {Cambridge University Press},\ \bibinfo {year}
  {2011})\BibitemShut {NoStop}%
\bibitem [{\citenamefont {Dominicis}(1976)}]{dominicis1976technics}%
  \BibitemOpen
  \bibfield  {author} {\bibinfo {author} {\bibfnamefont {C.~d.}\ \bibnamefont
  {Dominicis}},\ }\bibfield  {title} {\bibinfo {title} {Technics of field
  renormalization and dynamics of critical phenomena},\ }in\ \href@noop {}
  {\emph {\bibinfo {booktitle} {J. Phys.(Paris), Colloq}}}\ (\bibinfo {year}
  {1976})\ pp.\ \bibinfo {pages} {C1--247}\BibitemShut {NoStop}%
\bibitem [{\citenamefont {Janssen}(1976)}]{janssen1976lagrangean}%
  \BibitemOpen
  \bibfield  {author} {\bibinfo {author} {\bibfnamefont {H.-K.}\ \bibnamefont
  {Janssen}},\ }\bibfield  {title} {\bibinfo {title} {On a lagrangean for
  classical field dynamics and renormalization group calculations of dynamical
  critical properties},\ }\href@noop {} {\bibfield  {journal} {\bibinfo
  {journal} {Zeitschrift f{\"u}r Physik B Condensed Matter}\ }\textbf {\bibinfo
  {volume} {23}},\ \bibinfo {pages} {377} (\bibinfo {year} {1976})}\BibitemShut
  {NoStop}%
\bibitem [{SM()}]{SM}%
  \BibitemOpen
  \href@noop {} {\bibinfo  {journal} {Supplemental material}\ }\BibitemShut
  {NoStop}%
\bibitem [{\citenamefont {Wiese}(1998)}]{WieseSHE}%
  \BibitemOpen
\bibfield  {journal} {  }\bibfield  {author} {\bibinfo {author} {\bibfnamefont
  {K.~J.}\ \bibnamefont {Wiese}},\ }\bibfield  {title} {\bibinfo {title} {On
  the perturbation expansion of the kpz equation},\ }\href
  {https://doi.org/10.1023/B:JOSS.0000026730.76868.c4} {\bibfield  {journal}
  {\bibinfo  {journal} {Journal of Statistical Physics}\ }\textbf {\bibinfo
  {volume} {93}},\ \bibinfo {pages} {143} (\bibinfo {year} {1998})}\BibitemShut
  {NoStop}%
\bibitem [{\citenamefont {Canet}\ \emph {et~al.}(2010)\citenamefont {Canet},
  \citenamefont {Chat\'e}, \citenamefont {Delamotte},\ and\ \citenamefont
  {Wschebor}}]{nonperturbativeRGLeonieCanet}%
  \BibitemOpen
  \bibfield  {author} {\bibinfo {author} {\bibfnamefont {L.}~\bibnamefont
  {Canet}}, \bibinfo {author} {\bibfnamefont {H.}~\bibnamefont {Chat\'e}},
  \bibinfo {author} {\bibfnamefont {B.}~\bibnamefont {Delamotte}},\ and\
  \bibinfo {author} {\bibfnamefont {N.}~\bibnamefont {Wschebor}},\ }\bibfield
  {title} {\bibinfo {title} {Nonperturbative renormalization group for the
  kardar-parisi-zhang equation},\ }\href
  {https://doi.org/10.1103/PhysRevLett.104.150601} {\bibfield  {journal}
  {\bibinfo  {journal} {Phys. Rev. Lett.}\ }\textbf {\bibinfo {volume} {104}},\
  \bibinfo {pages} {150601} (\bibinfo {year} {2010})}\BibitemShut {NoStop}%
\bibitem [{\citenamefont {Mu}\ \emph {et~al.}(2024)\citenamefont {Mu},
  \citenamefont {Gong},\ and\ \citenamefont
  {Lemari\'e}}]{LemariePRLKPZAnderson}%
  \BibitemOpen
  \bibfield  {author} {\bibinfo {author} {\bibfnamefont {S.}~\bibnamefont
  {Mu}}, \bibinfo {author} {\bibfnamefont {J.}~\bibnamefont {Gong}},\ and\
  \bibinfo {author} {\bibfnamefont {G.}~\bibnamefont {Lemari\'e}},\ }\bibfield
  {title} {\bibinfo {title} {Kardar-parisi-zhang physics in the density
  fluctuations of localized two-dimensional wave packets},\ }\href
  {https://doi.org/10.1103/PhysRevLett.132.046301} {\bibfield  {journal}
  {\bibinfo  {journal} {Phys. Rev. Lett.}\ }\textbf {\bibinfo {volume} {132}},\
  \bibinfo {pages} {046301} (\bibinfo {year} {2024})}\BibitemShut {NoStop}%
\bibitem [{\citenamefont
  {{\v{Z}}nidari{\v{c}}}(2010{\natexlab{a}})}]{Znidaric__dephasingXXZ}%
  \BibitemOpen
  \bibfield  {author} {\bibinfo {author} {\bibfnamefont {M.}~\bibnamefont
  {{\v{Z}}nidari{\v{c}}}},\ }\bibfield  {title} {\bibinfo {title}
  {Dephasing-induced diffusive transport in the anisotropic heisenberg model},\
  }\href {https://doi.org/10.1088/1367-2630/12/4/043001} {\bibfield  {journal}
  {\bibinfo  {journal} {New Journal of Physics}\ }\textbf {\bibinfo {volume}
  {12}},\ \bibinfo {pages} {043001} (\bibinfo {year}
  {2010}{\natexlab{a}})}\BibitemShut {NoStop}%
\bibitem [{\citenamefont
  {{\v{Z}}nidari{\v{c}}}(2010{\natexlab{b}})}]{Znidaric__XXdeph}%
  \BibitemOpen
  \bibfield  {author} {\bibinfo {author} {\bibfnamefont {M.}~\bibnamefont
  {{\v{Z}}nidari{\v{c}}}},\ }\bibfield  {title} {\bibinfo {title} {Exact
  solution for a diffusive nonequilibrium steady state of an open quantum
  chain},\ }\href {https://doi.org/10.1088/1742-5468/2010/05/l05002} {\bibfield
   {journal} {\bibinfo  {journal} {Journal of Statistical Mechanics: Theory and
  Experiment}\ }\textbf {\bibinfo {volume} {2010}},\ \bibinfo {pages} {L05002}
  (\bibinfo {year} {2010}{\natexlab{b}})}\BibitemShut {NoStop}%
\bibitem [{\citenamefont {{\v{Z}}nidari{\v{c}}}\ and\ \citenamefont
  {Horvat}(2013)}]{Znidaric_dephasing}%
  \BibitemOpen
  \bibfield  {author} {\bibinfo {author} {\bibfnamefont {M.}~\bibnamefont
  {{\v{Z}}nidari{\v{c}}}}\ and\ \bibinfo {author} {\bibfnamefont
  {M.}~\bibnamefont {Horvat}},\ }\bibfield  {title} {\bibinfo {title}
  {Transport in a disordered tight-binding chain with dephasing},\ }\href
  {https://doi.org/10.1140/epjb/e2012-30730-9} {\bibfield  {journal} {\bibinfo
  {journal} {The European Physical Journal B}\ }\textbf {\bibinfo {volume}
  {86}},\ \bibinfo {pages} {67} (\bibinfo {year} {2013})}\BibitemShut {NoStop}%
\bibitem [{\citenamefont {Bauer}\ \emph {et~al.}(2017)\citenamefont {Bauer},
  \citenamefont {Bernard},\ and\ \citenamefont
  {Jin}}]{BauerBernardJinScipostfluctuatinghydro}%
  \BibitemOpen
  \bibfield  {author} {\bibinfo {author} {\bibfnamefont {M.}~\bibnamefont
  {Bauer}}, \bibinfo {author} {\bibfnamefont {D.}~\bibnamefont {Bernard}},\
  and\ \bibinfo {author} {\bibfnamefont {T.}~\bibnamefont {Jin}},\ }\bibfield
  {title} {\bibinfo {title} {{Stochastic dissipative quantum spin chains (I) :
  Quantum fluctuating discrete hydrodynamics}},\ }\href
  {https://doi.org/10.21468/SciPostPhys.3.5.033} {\bibfield  {journal}
  {\bibinfo  {journal} {SciPost Phys.}\ }\textbf {\bibinfo {volume} {3}},\
  \bibinfo {pages} {033} (\bibinfo {year} {2017})}\BibitemShut {NoStop}%
\bibitem [{\citenamefont {Bastianello}\ \emph {et~al.}(2020)\citenamefont
  {Bastianello}, \citenamefont {De~Nardis},\ and\ \citenamefont
  {De~Luca}}]{BastianelloDeNardisDeluca_GHDDeph}%
  \BibitemOpen
  \bibfield  {author} {\bibinfo {author} {\bibfnamefont {A.}~\bibnamefont
  {Bastianello}}, \bibinfo {author} {\bibfnamefont {J.}~\bibnamefont
  {De~Nardis}},\ and\ \bibinfo {author} {\bibfnamefont {A.}~\bibnamefont
  {De~Luca}},\ }\bibfield  {title} {\bibinfo {title} {Generalized hydrodynamics
  with dephasing noise},\ }\href {https://doi.org/10.1103/PhysRevB.102.161110}
  {\bibfield  {journal} {\bibinfo  {journal} {Phys. Rev. B}\ }\textbf {\bibinfo
  {volume} {102}},\ \bibinfo {pages} {161110} (\bibinfo {year}
  {2020})}\BibitemShut {NoStop}%
\bibitem [{\citenamefont {Noel}\ \emph {et~al.}(2022)\citenamefont {Noel},
  \citenamefont {Niroula}, \citenamefont {Zhu}, \citenamefont {Risinger},
  \citenamefont {Egan}, \citenamefont {Biswas}, \citenamefont {Cetina},
  \citenamefont {Gorshkov}, \citenamefont {Gullans}, \citenamefont {Huse},\
  and\ \citenamefont {Monroe}}]{NaturePhyMiPTExp}%
  \BibitemOpen
  \bibfield  {author} {\bibinfo {author} {\bibfnamefont {C.}~\bibnamefont
  {Noel}}, \bibinfo {author} {\bibfnamefont {P.}~\bibnamefont {Niroula}},
  \bibinfo {author} {\bibfnamefont {D.}~\bibnamefont {Zhu}}, \bibinfo {author}
  {\bibfnamefont {A.}~\bibnamefont {Risinger}}, \bibinfo {author}
  {\bibfnamefont {L.}~\bibnamefont {Egan}}, \bibinfo {author} {\bibfnamefont
  {D.}~\bibnamefont {Biswas}}, \bibinfo {author} {\bibfnamefont
  {M.}~\bibnamefont {Cetina}}, \bibinfo {author} {\bibfnamefont {A.~V.}\
  \bibnamefont {Gorshkov}}, \bibinfo {author} {\bibfnamefont {M.~J.}\
  \bibnamefont {Gullans}}, \bibinfo {author} {\bibfnamefont {D.~A.}\
  \bibnamefont {Huse}},\ and\ \bibinfo {author} {\bibfnamefont
  {C.}~\bibnamefont {Monroe}},\ }\bibfield  {title} {\bibinfo {title}
  {Measurement-induced quantum phases realized in a trapped-ion quantum
  computer},\ }\href {https://doi.org/10.1038/s41567-022-01619-7} {\bibfield
  {journal} {\bibinfo  {journal} {Nature Physics}\ }\textbf {\bibinfo {volume}
  {18}},\ \bibinfo {pages} {760} (\bibinfo {year} {2022})}\BibitemShut
  {NoStop}%
\bibitem [{\citenamefont {Koh}\ \emph {et~al.}(2023)\citenamefont {Koh},
  \citenamefont {Sun}, \citenamefont {Motta},\ and\ \citenamefont
  {Minnich}}]{NatureMiPTexperiment}%
  \BibitemOpen
  \bibfield  {author} {\bibinfo {author} {\bibfnamefont {J.~M.}\ \bibnamefont
  {Koh}}, \bibinfo {author} {\bibfnamefont {S.-N.}\ \bibnamefont {Sun}},
  \bibinfo {author} {\bibfnamefont {M.}~\bibnamefont {Motta}},\ and\ \bibinfo
  {author} {\bibfnamefont {A.~J.}\ \bibnamefont {Minnich}},\ }\bibfield
  {title} {\bibinfo {title} {Measurement-induced entanglement phase transition
  on a superconducting quantum processor with mid-circuit readout},\ }\bibfield
   {journal} {\bibinfo  {journal} {Nature Physics}\ }\href
  {https://doi.org/10.1038/s41567-023-02076-6} {10.1038/s41567-023-02076-6}
  (\bibinfo {year} {2023})\BibitemShut {NoStop}%
\bibitem [{\citenamefont {{Poboiko}}\ \emph
  {et~al.}(2023{\natexlab{b}})\citenamefont {{Poboiko}}, \citenamefont
  {{Gornyi}},\ and\ \citenamefont {{Mirlin}}}]{MirlinGornyiMiPT3DFreefermions}%
  \BibitemOpen
  \bibfield  {author} {\bibinfo {author} {\bibfnamefont {I.}~\bibnamefont
  {{Poboiko}}}, \bibinfo {author} {\bibfnamefont {I.~V.}\ \bibnamefont
  {{Gornyi}}},\ and\ \bibinfo {author} {\bibfnamefont {A.~D.}\ \bibnamefont
  {{Mirlin}}},\ }\bibfield  {title} {\bibinfo {title} {{Measurement-induced
  phase transition for free fermions above one dimension}},\ }\href@noop {}
  {\bibfield  {journal} {\bibinfo  {journal} {arXiv e-prints}\ ,\ \bibinfo
  {eid} {arXiv:2309.12405}} (\bibinfo {year} {2023}{\natexlab{b}})},\ \Eprint
  {https://arxiv.org/abs/2309.12405} {arXiv:2309.12405 [quant-ph]} \BibitemShut
  {NoStop}%
\bibitem [{\citenamefont {{Chahine}}\ and\ \citenamefont
  {{Buchhold}}(2023)}]{BuchholdMiPT2d}%
  \BibitemOpen
  \bibfield  {author} {\bibinfo {author} {\bibfnamefont {K.}~\bibnamefont
  {{Chahine}}}\ and\ \bibinfo {author} {\bibfnamefont {M.}~\bibnamefont
  {{Buchhold}}},\ }\bibfield  {title} {\bibinfo {title} {{Entanglement phases,
  localization and multifractality of monitored free fermions in two
  dimensions}},\ }\href {https://doi.org/10.48550/arXiv.2309.12391} {\bibfield
  {journal} {\bibinfo  {journal} {arXiv e-prints}\ ,\ \bibinfo {eid}
  {arXiv:2309.12391}} (\bibinfo {year} {2023})},\ \Eprint
  {https://arxiv.org/abs/2309.12391} {arXiv:2309.12391 [cond-mat.str-el]}
  \BibitemShut {NoStop}%
\bibitem [{\citenamefont {Mannella}(1997)}]{mannella1997numerical}%
  \BibitemOpen
  \bibfield  {author} {\bibinfo {author} {\bibfnamefont {R.}~\bibnamefont
  {Mannella}},\ }\bibfield  {title} {\bibinfo {title} {Numerical integration of
  stochastic differential equations},\ }\href@noop {} {\bibfield  {journal}
  {\bibinfo  {journal} {arXiv preprint cond-mat/9709326}\ } (\bibinfo {year}
  {1997})}\BibitemShut {NoStop}%
\bibitem [{\citenamefont {Moss}\ and\ \citenamefont
  {McClintock}(1989)}]{moss_experiments_1989}%
  \BibitemOpen
  \bibinfo {editor} {\bibfnamefont {F.}~\bibnamefont {Moss}}\ and\ \bibinfo
  {editor} {\bibfnamefont {P.~V.~E.}\ \bibnamefont {McClintock}},\ eds.,\
  \href@noop {} {\emph {\bibinfo {title} {Experiments and simulations}}},\
  \bibinfo {series} {Noise in nonlinear dynamical systems}\ No.\ \bibinfo
  {number} {v. 3}\ (\bibinfo  {publisher} {Cambridge University Press},\
  \bibinfo {address} {Cambridge [Cambridgeshire] ; New York},\ \bibinfo {year}
  {1989})\BibitemShut {NoStop}%
\end{thebibliography}%

\appendix
\onecolumngrid
\section{Martin-Siggia-Rose-Janssen-De Dominicis (MSRJD) action for complex fields}
\label{sec:Martin-Siggia-Rose-(MSR)-action}

In this section we derive the MSRJD action of the stochastic differential
equation (SDE) : 
\begin{equation}
\label{eq:local_equation}
d\varphi=\left(iD\nabla^{2}\varphi-\frac{\gamma}{2}\varphi\right)dt+\sqrt{\lambda}\varphi d\eta.
\end{equation}
Since the field $\varphi$ is complex, we need to treat individually
the real and imaginary part. Defining $X:=\Re(\varphi)$ and $Y:=\Im(\varphi)$, we obtain the coupled equations 
\begin{align}
dX & =\left(-D\nabla^{2}Y-\frac{\gamma}{2}X\right)dt+\sqrt{\lambda}Xd\eta, &
dY & =\left(D\nabla^{2}X-\frac{\gamma}{2}Y\right)dt+\sqrt{\lambda}Yd\eta.
\end{align}
The partition function is then defined as 
\begin{equation}
Z:=\mathbb{E}\left[\int{\cal D}[X,Y]\delta\left(\partial_{t}X+D\nabla^{2}Y+\frac{\gamma}{2}X-\sqrt{\lambda}X\eta\right)\delta\left(\partial_{t}Y-D\nabla^{2}X+\frac{\gamma}{2}Y-\sqrt{\lambda}Y\eta\right)\right],
\end{equation}
where $\mathbb{E}[\cdot]$ indicates an average with respect to the noise $\eta$. We introduce the identity resolution at every point in space-time
\begin{equation}
\delta\left(f(X)\right):=\int{\cal D}[X^{a},Y^{a}]e^{-i\left(X^{a}+Y^{a}\right)f(X)}\;,
\end{equation}
where $f(X)$ is an arbitrary test function. The partition function then reads
\begin{equation}
Z=\mathbb{E}\left[\int{\cal D}[X,X^{a},Y,Y^{a}]e^{-i\int drdtX^{a}\left(\partial_{t}X+D\nabla^{2}Y+\frac{\gamma}{2}X-\sqrt{\lambda}X\eta\right)}e^{-i\int drdtY^{a}\left(\partial_{t}Y-D\nabla^{2}X+\frac{\gamma}{2}Y-\sqrt{\lambda}Y\eta\right)}\right].
\end{equation}
Further introducing the fields 
% \begin{align}
% \varphi :=& X+iY,\quad\overline{\varphi}:=& X-iY, \varphi^{a} :=& X^{a}+iY^{a},\quad\overline{\varphi}^{a}:=& X^{a}-iY^{a},
% \end{align}
\begin{align}
\varphi :=& X+iY, & \overline{\varphi}:=& X-iY, & \varphi^{a} :=& X^{a}+iY^{a}, & \overline{\varphi}^{a}:=& X^{a}-iY^{a},
\end{align}
the partition function can be recasted into
\begin{equation}
\label{eq:Z_phi}
Z=\mathbb{E}\left[\int{\cal D}[\varphi,\varphi^{a},\overline{\varphi},\overline{\varphi}^{a}]e^{\frac{i}{2}\int d^{d}rdt\left[iD\left(\overline{\varphi}^{a}\nabla^{2}\varphi-\varphi^{a}\nabla^{2}\overline{\varphi}\right)-\left(\varphi^{q}\partial_{t}\overline{\varphi}+\overline{\varphi}^{a}\partial_{t}\varphi\right)-\frac{\gamma}{2}\left(\overline{\varphi}\varphi^{a}+\varphi\overline{\varphi}^{a}\right)+\sqrt{\lambda}\eta_{t}\left(\varphi^{a}\overline{\varphi}+\overline{\varphi}^{a}\varphi\right)\right]}\right].
\end{equation}
For later convenience, we recall the implicit discrete time version
of \eqref{eq:Z_phi} imposed by the It\=o convention. 
Denoting $\varphi_{n}=\varphi\left(n\Delta t\right)$, this discrete time version reads
\begin{align}
Z & =\mathbb{E}\bigg[\int{\cal D}[\varphi,\varphi^{a},\overline{\varphi},\overline{\varphi}^{a}]\\
 & e^{\frac{i}{2}\int d^{d}r\sum_{n}\left[iD\left(\overline{\varphi}_{n+1}^{a}\nabla^{2}\varphi_{n}-\varphi_{n+1}^{a}\nabla^{2}\overline{\varphi}_{n}\right)-\left(\varphi_{n+1}^{q}\frac{\overline{\varphi}_{n+1}-\overline{\varphi}_{n}}{\Delta t}+\overline{\varphi}_{n+1}^{a}\frac{\varphi_{n+1}-\varphi_{n}}{\Delta t}\right)-\frac{\gamma}{2}\left(\overline{\varphi}_{n}\varphi_{n+1}^{a}+\varphi_{n}\overline{\varphi}_{n+1}^{a}\right)+\sqrt{\lambda}\eta_{n}\left(\varphi_{n+1}^{a}\overline{\varphi}_{n}+\overline{\varphi}_{n+1}^{a}\varphi_{n}\right)\right]}\bigg].\nonumber 
\end{align}
Note that, importantly, the auxiliary fields are always evaluated
one step ahead of the physical fields in the noise term. We now perform the average over the Gaussian noise using the formula
\begin{equation}
\mathbb{E}[f(\eta)]=\int{\cal D}[\eta]e^{-\int drdt\frac{\eta^{2}}{2}}f(\eta)
\end{equation}
to get 
\begin{align}
Z & =\int{\cal D}[\varphi,\varphi^{a},\overline{\varphi},\overline{\varphi}^{a}]e^{iS},
\end{align}
where the action $S=S_{0}+S_{\nu}$ contains two contributions respectively given by 
\begin{comment}
\begin{equation}
S=\int drdt\left[-X^{a}\left(\partial_{t}X+D\nabla^{2}Y+\frac{\gamma}{2}X\right)-Y^{a}\left(\partial_{t}Y-D\nabla^{2}X+\frac{\gamma}{2}Y\right)+i\frac{\lambda}{2}\left(XX^{a}+YY^{a}\right)^{2}\right].
\end{equation}

Reintroducing the complex form of the fields $\varphi=X+iY$ and $\bar{\varphi}=X-iY$, we get
\end{comment}
\begin{align}
S_{0} & =\frac{1}{2}\int d^{d}\vec{r}dt(\bar{\varphi},\bar{\varphi}^{a})G_{0}^{-1}\begin{pmatrix}\varphi\\
\varphi^{a}
\end{pmatrix}, &
% G_{0}^{-1} & =\begin{pmatrix}0 & -\frac{\gamma}{2}+\partial_{t}^{A}-iD\nabla^{2}\\
% -\frac{\gamma}{2}-\partial_{t}^{R}+iD\nabla^{2} & 0
% \end{pmatrix},\\
S_{\nu} & =\frac{i}{8}\int d^{d}\vec{r}dt\big(\lambda^{{\rm I}}\left(\bar{\varphi}^{a}\right)^{2}\varphi^{2}+2\lambda^{{\rm II}}\bar{\varphi}^{a}\bar{\varphi}\varphi^{a}\varphi+\lambda^{{\rm I}*}\bar{\varphi}^{2}\left(\varphi^{a}\right)^{2}\big)\;,
\label{eq:snuapp}
\end{align}
for which the quadratic matrix $G_{0}^{-1}$ reads
\begin{align}
G_{0}^{-1} & =\begin{pmatrix}0 & -\frac{\gamma}{2}+\partial_{t}^{A}-iD\nabla^{2}\\
-\frac{\gamma}{2}-\partial_{t}^{R}+iD\nabla^{2} & 0
\end{pmatrix}\;.
\end{align}
In \eqref{eq:snuapp}, we give distinct names for the different $\lambda$ coefficients as they will follow distinct RG flows. 
Remark that the $4$-vertex interactions always contain two quantum and two auxiliary fields as well as the same number of conjugated and non-conjugated fields. 
The retarded and advanced components in momentum and frequency space
are obtained from inverting $G_{0}^{-1}$ as 
\begin{align}
G_{0}^{R}(\vec{q},\omega) & :=-i\int d^{d}\vec{r}dte^{i(\omega t-\vec{q}.\vec{r})}\langle \varphi(r,t)\bar{\varphi}^{a}(0,0)\rangle_{S_{0}}=\frac{-2i}{\omega-Dq^{2}+i\frac{\gamma}{2}},\\
G_{0}^{A}(\vec{q},\omega) & :=-i\int d^{d}\vec{r}dte^{i(\omega t-\vec{q}.\vec{r})}\langle \varphi^{a}(r,t)\bar{\varphi}(0,0)\rangle_{S_{0}}=\frac{2i}{\omega-Dq^{2}-i\frac{\gamma}{2}},
\end{align}
where $\langle\rangle_{S_{0}}$ indicates averaging over the gaussian action $S_{0}$.

\section{One-loop perturbative renormalization}
\label{sec:One-loop-perturbative-renormaliz}

In this section we provide details on the renormalization 
%of the mass term $\gamma$ and 
of the interaction term $\lambda$.
We use the momentum-shell cut-off procedure and keep the perturbation
in $\lambda$ at one-loop level. We call $\Lambda$ the momentum cut-off
and $z$, $\chi$, $\bar{\chi}$ the dynamical exponents
associated to $t$, $\varphi$, $\bar{\varphi}$ respectively. The renormalization
flow is parametrized by the scale $l$. 

\begin{figure}
\includegraphics[width=0.87\textwidth]{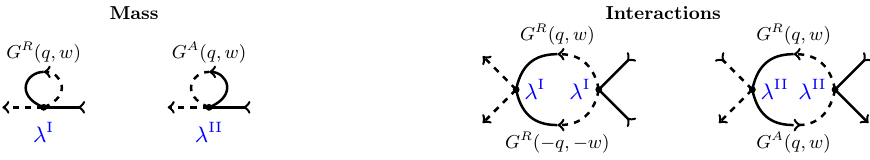}
\caption{One-loop contributions to the RG of the mass and interactions. Dashed lines indicate the auxiliary field indexed by $a$.
%\tj{Do we need the figure as it is already in the main? Maybe we can get rid of the one in the main because we do not cite it?} 
\label{fig:One-loop-contributions-toAPP}}
\end{figure}

As explained in the main text, at one-loop level, there are no
diagrams renormalizing the terms proportional to $\partial_{t}$
and $\nabla^{2}$ in the action. Enforcing stationarity in the RG flow for the corresponding
prefactors gives 
\begin{equation}
\chi+\bar{\chi}+d=0,\quad z=2.\label{eq:dynamicalexponentstrivial}
\end{equation}
The Feynman diagrams contributing to the renormalization of $\gamma$
and $\lambda$ are shown graphically on Fig.~\ref{fig:One-loop-contributions-toAPP} 
In the I\=o convention, as mentionned above, the auxiliary fields of the interaction vertex are always evaluated one step ahead of the physical fields. Thus, the tadpoles diagram renormalizing the mass have to vanish due to causality.
Therefore, the renormalization of the mass term is obtained by simple dimensional analysis as 
\begin{equation}
\gamma=\gamma_0 e^{2l}.
\end{equation}

Before going on with the renormalization of $\lambda$, we begin by providing some useful identities for the loop integrals. 
% \textbf{\textit{Abacus for the one-loop integrals.}} 
As usual in the RG procedure, we separate the fields $\varphi$
and $\bar{\varphi}$ into fast momenta $\Lambda>q>\Lambda e^{-\delta l}$
denoted with an $f$ and slow momenta $\Lambda e^{-\delta l}>q$ denoted
with an $s$. We compute here all the fast momenta loop integrals that show up in the renormalization %of $\gamma$ and 
of $\lambda^{{\rm I/II}}$.

Furthermore, the two-propagators loop renormalizing $\lambda^{\rm I}$ is given by 

\begin{align}
\nonumber
\int_{f}\frac{d^{{\rm d}}\vec{q}d\omega}{(2\pi)^{({\rm d}+1)}}G^{R}(\vec{q},\omega)G^{R}(-\vec{q},-\omega) =4\int_{f}\frac{d^{{\rm d}}\vec{q}d\omega}{(2\pi)^{({\rm d}+1)}}\frac{1}{\left(\omega-Dq^{2}+i\frac{\gamma}{2}\right)\left(\omega+Dq^{2}-i\frac{\gamma}{2}\right)}=& 4i\int_{f}\frac{d^{{\rm d}}\vec{q}}{(2\pi)^{{\rm d}}}\frac{1}{-2Dq^{2}+i\gamma} \\=&4K_{{\rm d}}\delta l\frac{1}{\gamma+i2D\Lambda^{2}},\label{eq:GRGRloop}
\end{align}
and its complex conjugate  
\begin{equation}
\int_{f}\frac{d^{{\rm d}}\vec{q}d\omega}{(2\pi)^{({\rm d}+1)}}G^{A}(\vec{q},\omega)G^{A}(-\vec{q},-\omega)=4K_{{\rm d}}\delta l\frac{1}{\gamma-i2D\Lambda^{2}}.
\end{equation}

We now turn to the derivation of the flow equations for the interaction terms.
After one step of renormalization, $\lambda^{{\rm I}}$ is renormalized into $\lambda^{{\rm I}'}$ which is given by
\begin{align}
\lambda^{{\rm I}'} & =e^{\delta l\left(2\chi+2\bar{\chi}+{\rm d}+z\right)}\left(\lambda^{{\rm I}}+\frac{1}{4}\left(\lambda^{{\rm I}}\right)^{2}\int_{f}\frac{d^{{\rm d}}\vec{q}d\omega}{(2\pi)^{{\rm d}+1}}G^{R}(\vec{q},\omega)G^{R}(-\vec{q},-\omega)\right),
\label{eq:renorm_lambda_1}
\end{align}
For an infinitesimal step $\delta l$, \eqref{eq:renorm_lambda_1} reads
\begin{equation}
\frac{d\lambda^{{\rm I}}}{dl}=(2-{\rm d})\lambda^{{\rm I}}+K_{{\rm d}}\frac{\left(\lambda^{{\rm I}}\right)^{2}}{\gamma+i2D\Lambda^{2}},
\end{equation}
where we made use of Eqs. \eqref{eq:dynamicalexponentstrivial} and \eqref{eq:GRGRloop}. 
Doing a similar computation for $\lambda^{{\rm II}}$, one has %
\begin{align}
\lambda^{{\rm II}'} & =e^{\delta l\left(2\chi+2\bar{\chi}+{\rm d}+z\right)}\left(\lambda^{{\rm II}}+\frac{1}{4}\left(\lambda^{{\rm II}}\right)^{2}\int_{f}\frac{d^{{\rm d}}\vec{q}d\omega}{(2\pi)^{{\rm d}+1}}|G^{R}(\vec{q},\omega)|^{2}\right),
\end{align}
which, for an infinitesimal step, becomes
\begin{equation}
\frac{d\lambda^{{\rm II}}}{dl}=(2-{\rm d})\lambda^{{\rm II}}+\frac{K_{{\rm d}}}{\gamma}\left(\lambda^{{\rm II}}\right)^{2}.
\end{equation}
For convenience, we also provide the flow equations for the real and imaginary parts of $\lambda^{\rm I}$ as 
\begin{align}
\frac{d\lambda_{R}^{{\rm I}}}{dl}= & (2-{\rm d})\lambda_{R}^{{\rm I}}+K_{{\rm d}}\left(\frac{\left(\left(\lambda_{R}^{{\rm I}}\right)^{2}-\left(\lambda_{I}^{{\rm I}}\right)^{2}\right)\gamma+2D_{e}\lambda_{R}^{{\rm I}}\lambda_{I}^{{\rm I}}}{\gamma^{2}+D_{e}^{2}}\right),\\
\frac{d\lambda_{I}^{{\rm I}}}{dl}= & (2-{\rm d})\lambda_{I}^{{\rm I}}+K_{{\rm d}}\left(\frac{\left(\left(\lambda_{I}^{{\rm I}}\right)^{2}-\left(\lambda_{R}^{{\rm I}}\right)^{2}\right)D_{e}+2\lambda_{R}^{{\rm I}}\lambda_{I}^{{\rm I}}\gamma}{\gamma^{2}+D_{e}^{2}}\right).
\end{align}

\begin{figure}
\begin{tikzpicture}
% \tikzmath{\legy=-1.5;\scaleleg=1.3;}
% \node[font=\bf,scale=\scaleleg] at (-3.2,\legy) {a.};
% \node[font=\bf,scale=\scaleleg] at (2.5,\legy) {b.};
% \node[font=\bf,scale=\scaleleg] at (6,\legy) {c.};
\node at (0,0) {\includegraphics[width=\textwidth]{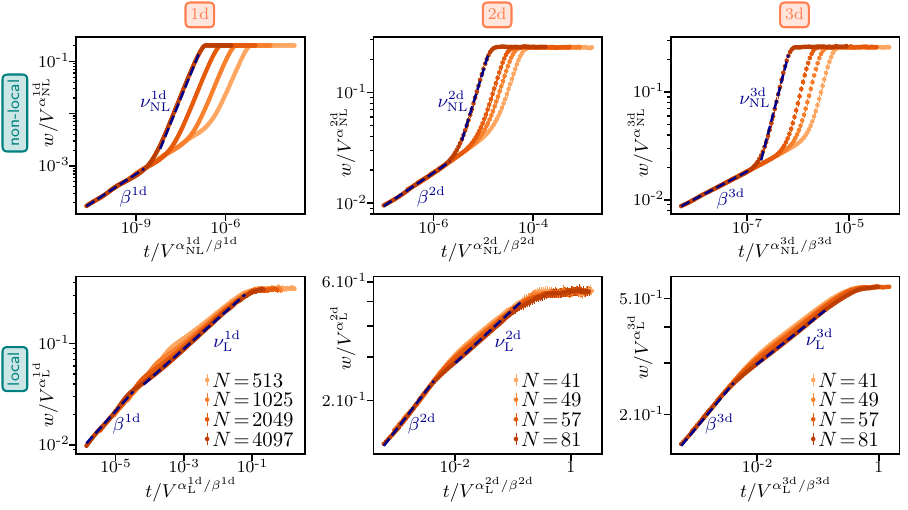}};
\end{tikzpicture}
\caption{\textbf{a.} Log-Log plot of the rescaled width as a function of the rescaled time for \eqref{eq:fundamentalequationpositionbasis} (first row) and \eqref{eq:local_equation} (second row) in different dimensions (columns) within the rough phase. 
Exponents: $\alpha^{1\rm{d}}_{\rm{NL}}=1$, $\nu^{1\rm{d}}_{\rm{NL}}=1.41$, $\alpha^{2\rm{d}}_{\rm{NL}}=0.48$, $\nu^{2\rm{d}}_{\rm{NL}}=1.46$, $\alpha^{3\rm{d}}_{\rm{NL}}=0.32$, $\nu^{3\rm{d}}_{\rm{NL}}=1.83$, $\alpha^{1\rm{d}}_{\rm{L}}=0.5$, $\nu^{1\rm{d}}_{\rm{L}}=0.30$, $\alpha^{2\rm{d}}_{\rm{L}}=0.18$, $\nu^{2\rm{d}}_{\rm{L}}=0.21$, $\alpha^{3\rm{d}}_{\rm{L}}=0.09$, $\nu^{3\rm{d}}_{\rm{L}}=0.16$, $\beta^{1\rm{d}}=0.37$, $\beta^{2\rm{d}}=0.30$, $\beta^{3\rm{d}}=0.25$.
\textbf{Parameters}: in $1\rm{d}$ we took $dt=0.001$, $\gamma=1$ for the nonlocal case and $\lambda=\gamma=2$ for the local case. 
In $2\rm{d}$ we took $dt=0.01$ and $\lambda=\gamma=5$.
In $3\rm{d}$, $dt=0.01$ and $\lambda=\gamma=8$.
$\tau =1$ in all dimensions.
}
\end{figure}

\section{Details on the numerics}
We recall the microscopic evolution of the quantum walker
\begin{align}
d\psi_{\boldsymbol{j}}= & i\tau\sum_{\{|\boldsymbol{e}|=1\}}\psi_{\boldsymbol{j}+\boldsymbol{e}}dt-\frac{\gamma}{2}\psi_{\boldsymbol{j}}\big(1-2|\psi_{\boldsymbol{j}}|^{2}+\sum_{\boldsymbol{m}}|\psi_{\boldsymbol{m}}|^{4}\big)dt +\sqrt{\gamma}\psi_{\boldsymbol{j}}\big(dB_{t}^{\boldsymbol{j}}-\sum_{\boldsymbol{m}}|\psi_{\boldsymbol{m}}|^{2}dB_{t}^{\boldsymbol{m}}\big),\label{eq:fundamentalequationpositionbasis_app}
\end{align}
as well as the definition of the width along with its Family-Vicsek scaling
\begin{equation}
w:=\sqrt{\frac{1}{V}\sum_{\boldsymbol{j}} (h_{\boldsymbol{j}}-\langle h\rangle_{\rm s})^2}\propto V^\alpha f\left(\frac{t}{V^{\alpha/\beta}}\right).
\label{eq:family_vicsek_app}
\end{equation}
In this appendix, we describe the methods employed to numerically integrate \eqref{eq:fundamentalequationpositionbasis_app} from the main text.
We first note the peculiar structure of the noise term $dB_{t}^{\boldsymbol{j}}$ in \eqref{eq:fundamentalequationpositionbasis_app}: it is a multiplicative multi-dimensional noise.
Thus, \eqref{eq:fundamentalequationpositionbasis_app} falls into the class of stochastic differential equations taking the form
\begin{equation}
    \label{eq:multidimensional_SDE}
    \frac{dx_{\boldsymbol{j}}}{dt} = f_{\boldsymbol{j}}(\{\bx\}) + \sum_{\boldsymbol{i}} g_{\boldsymbol{ji}}(\{\bx\})\xi_{\boldsymbol{i}}
\end{equation}
where $\xi_{\boldsymbol{i}}$'s are gaussian white noises such that $\langle\xi_{\boldsymbol{i}}(s)\xi_{\boldsymbol{k}}(s')\rangle=\delta_{\boldsymbol{ik}}\delta(s-s')$, and $f_{\boldsymbol{j}}$ and $g_{\boldsymbol{ji}}$ are functions of the set of position $\{\bx\}=\{x_{\boldsymbol{k}},\ \boldsymbol{k}\in [1,..,N^{\rm{d}}]\}$.
Numerical integration schemes for SDEs of type \eqref{eq:multidimensional_SDE} have been discussed in \cite{mannella1997numerical} and chapter 7 of \cite{moss_experiments_1989}.
The combination of the multiplicative and multi-dimensional nature of the noise in \eqref{eq:multidimensional_SDE} renders usual higher order Runge-Kutta-based SDE algorithms inoperative.
As described in \cite{moss_experiments_1989}, the two numerical schemes available for integrating \eqref{eq:multidimensional_SDE} are both of order $dt$ at maximum.
The first one is an Euler-Maruyama scheme (\textit{i.e.} simple forward Euler), which allows for a straightforward integration of \eqref{eq:multidimensional_SDE} in It\=o prescription.
The second one is a first order Runge-Kutta scheme with an approximate closure valid up to $dt$: it allows for numerical integration of \eqref{eq:multidimensional_SDE} directly in Stratonovitch prescription.
As we studied \eqref{eq:fundamentalequationpositionbasis_app} within It\=o formalism in the main text, we naturally choose the former Euler-Maruyama algorithm to perform our numerical integrations.
In addition, to ensure the correct normalization of the field $\psi_{\boldsymbol{j}}$ at each time (\textit{i.e.} that $\sum_{\boldsymbol{j}}|\psi_{\boldsymbol{j}}|^2=1$), we projected as $\psi_{\boldsymbol{j}}\to \psi_{\boldsymbol{j}}\big/\sqrt{\sum_{\boldsymbol{j}}|\psi_{\boldsymbol{j}}|^2}$ after each step forward in time.
To check the convergence of the algorithm, we divided the time step $dt$ by two and verified the stability of our results.
\\
The width $w$ at fixed $\gamma$ and fixed system size $N$ was obtained by averaging over at least 500 realizations of \eqref{eq:fundamentalequationpositionbasis_app}.
We made sure that simulations ran long enough for $w$ to effectively reach its plateau value at large time.
To compute the roughening exponent $\alpha$ at fixed $\gamma$, we performed a linear fit of the width's plateau value $w(t=\infty)$ as a function of the system size $N$ in Log-Log.
From \eqref{eq:family_vicsek_app}, we indeed have that $\log(w(t=\infty))\sim\alpha \log(N)$: the coefficient of the later linear fit gives $\alpha$.

\end{document}